\documentstyle[12pt,epsf]{article}
\def\href#1#2{#2}   
\newif\ifdraft
\draftfalse	  
\reversemarginpar   
\makeatletter
\let\mlabel=\label
\let\adkendequation=\endequation%
\def\endequation{\adkendequation\adklabel\global\@ignoretrue}
\let\adkendeqnarray=\endeqnarray%
\def\endeqnarray{\adkendeqnarray\adklabel\global\@ignoretrue}
\newbox\marglabbox
\def\adklabel{\ifvoid\marglabbox\else\marginpar{\unhbox\marglabbox}\fi}
\def\label#1{\ifdraft\ifmmode%
  \global\setbox\marglabbox=\hbox{\hfill\fbox{\tiny\verb*~#1~}}%
  \else\ifinner\else\marginpar{\hfill\fbox{\tiny\verb*~#1~}}%
  \fi\fi\fi \mlabel{#1}}
\makeatother
\ifdraft%
\fi
\font\twelvebb=msbm12
\font\tenbb=msbm10
\font\sevenbb=msbm7
  \newfam\bbfam
  
  \textfont\bbfam=\twelvebb
  \scriptfont\bbfam=\tenbb
  \scriptscriptfont\bbfam=\sevenbb
\font\twelveeusm=eusm10 scaled 1200
\font\teneusm=eusm10
  \newfam\eusmfam
  
  \textfont\eusmfam=\twelveeusm
  \scriptfont\eusmfam=\teneusm
  \scriptscriptfont\eusmfam=\scriptfont\eusmfam
\font\twelvefrak=eufm10 scaled 1200
\font\tenfrak=eufm10
  \newfam\frakfam
  
  \textfont\frakfam=\twelvefrak
  \scriptfont\frakfam=\tenfrak
  \scriptscriptfont\frakfam=\scriptfont\frakfam
\def\sqr#1#2{{\vcenter{\hrule height.#2pt
   \hbox{\vrule width.#2pt height#1pt \kern#1pt
      \vrule width.#2pt}
   \hrule height.#2pt}}}

\def\bsqr#1#2{{\vrule width #1pt height#2pt}}
\def\bsquare{{\mathchoice\bsqr66\bsqr66\bsqr33\bsqr33}}
\def\badbreak{\penalty1000}

\def\rational#1#2{{\mathchoice{\textstyle{#1\over#2}}%
  {\scriptstyle{#1\over#2}}{\scriptscriptstyle{#1\over#2}}{#1/#2}}}
\def\third{\rational13}			    
\newcommand{\muhat}{{\hat\mu}}              

\begin{document}

\begin{center}
{\Large{\bf Local Chirality of Low-Lying Dirac Eigenmodes}} \\
\vspace*{.1in}
{\Large{\bf and the Instanton Liquid Model}}\\
\vspace*{.4in}
{\large{I.~Horv\'ath$^1$,
S.J.~Dong$^1$,
T.~Draper$^1$, 
N.~Isgur$^3$,
F.X.~Lee$^{2,3}$,
K.F.~Liu$^1$}}\\
\vspace*{.1in}
{\large{
J.~McCune$^4$,
H.B.~Thacker$^4$ and
J.B.~Zhang$^5$}} \\
\vspace*{.15in}
$^1$University of Kentucky, Lexington, KY 40506\\
$^2$Center for Nuclear Studies and Department of Physics, 
George Washington University, Washington, DC 20052 \\
$^3$Jefferson Lab, 12000 Jefferson Avenue, Newport News, VA 23606 \\
$^4$Department of Physics, University of Virginia, Charlottesville, VA 22901 \\
$^5$CSSM and Department of Physics and Mathematical Physics, 
University of Adelaide, Adelaide, SA 5005, Australia

\vspace*{0.2in}
{\large{January 7 2002}}

\end{center}

\vspace*{0.15in}

\begin{abstract}
  \noindent
  The reasons for using low-lying Dirac eigenmodes to probe the local structure of 
  topological charge fluctuations in QCD are discussed, and it is pointed out that 
  the qualitative double-peaked behavior of the local chiral orientation probability 
  distribution in these modes is necessary, but not sufficient, for dominance of 
  instanton-like fluctuations. The results with overlap Dirac operator in Wilson gauge 
  backgrounds at lattice spacings ranging from $a\approx 0.04\,\mbox{\rm fm}$ to 
  $a\approx 0.12\,\mbox{\rm fm}$ are reported, and it is found that the size and density 
  of local structures responsible for double-peaking of the distribution are 
  in disagreement with the assumptions of the Instanton Liquid Model. More generally, 
  our results suggest that vacuum fluctuations of topological charge are not effectively
  dominated by locally quantized (integer-valued) lumps in QCD. 
\end{abstract}

\section{Introduction}

It has been recently suggested~\cite{Hor01A} that questions about the dynamical
nature of topological charge fluctuations in the QCD vacuum are worth revisiting. 
The notion that gauge field topology is relevant for understanding QCD started 
with the discovery of instantons~\cite{ins_disc}, and their subsequent use 
as a basis for, among other things, the qualitative resolution of the $U_A(1)$ 
problem~\cite{ins_tHooft}, and the discovery of $\Theta$-dependence of QCD 
physics~\cite{ins_theta}. These successes were associated with the use of 
semiclassical methods and a concrete {\it local} picture of the vacuum, 
the instanton gas picture~\cite{ins_gas}, characterized by formation of  
well-separated (anti)self-dual lumps of quantized topological charge. However, 
it was soon realized that such a vacuum does not lead to confinement, thus putting 
the relevance of instantons immediately into question. Reflecting upon this 
situation, Witten has argued~\cite{WittenUA(1)} that large quantum fluctuations 
associated with the confining vacuum naturally generate large fluctuations of topological 
charge, and lead to similar qualitative effects as those usually ascribed to instantons. 
In fact, he has conjectured that instantons are not important dynamically, arguing 
that the semiclassical picture is invalidated by large quantum corrections, and 
suggesting that topological charge fluctuates in a more or less continuous manner.

Nevertheless, QCD instantons remained rather popular as a frequently preferred 
(and analytically accessible) way of thinking about vacuum topology. 
The instanton solution was used as a basis for developing a rather successful
semiclassically-motivated phenomenology, the Instanton Liquid Model 
(ILM)~\cite{ILM}, where correlations among instantons were introduced to suppress
the infrared divergences present in the instanton gas. Even though termed 
a ``liquid'', the corresponding vacuum is quite dilute with (anti)instantons 
of radius $\rho\approx {\third}\,\mbox{\rm fm}$ and density 
$n\approx 1\,\mbox{\rm fm}^{-4}$ occupying a small fraction of space-time volume
and retaining their identity. This setup allows for an interesting mechanism 
of spontaneous chiral symmetry breaking~\cite{ins_mixing}, in which the t'Hooft 
``would-be'' zeromodes associated with individual (anti)instantons mix, and supply 
the finite density of near-zeromodes required by the Banks-Casher 
relation~\cite{Banks_Casher}. While the ILM is a phenomenological model, this 
elegant mixing picture invokes the impression that instantons play an important 
{\it microscopic} dynamical role in the QCD vacuum. A primary aim of our 
investigation is to examine whether the microscopic relevance of ILM picture
can be justified.

The support for the foundations of the ILM has frequently been drawn from lattice QCD
simulations, using equilibrated gauge configurations as typical representatives
in the path integral (see e.g. Ref.~\cite{ins_latsupport}). Indeed, it is quite 
reasonable to expect that lattice QCD will eventually provide us with detailed 
answers about the nature of topological charge fluctuations. However, finding 
a clean and satisfactory way to infer this information from lattice QCD 
has proven to be a nontrivial issue due to the fact that lattice gauge fields 
are found to be very rough at the scale of the lattice spacing. The apparent 
necessity to eliminate the short-distance fluctuations in some way resulted in 
manipulating the gauge fields in various cooling or smoothing procedures. While 
useful for estimating the global topological charge, these techniques are 
significantly biased as to the local structure of the vacuum they reveal, and 
there is always an inherent subjective element present. 

The idea of using fermions to study global topology is relatively 
old~\cite{fermion_glob}, but was not very much exploited due to both 
computational demands and problems with chiral symmetry. However, the  
suggestion to use low-lying fermionic modes to extract the information 
about {\it local} fluctuations of topological charge is recent~\cite{Hor01A} 
(see also Ref.~\cite{deGrand00A}). In the light of the above remarks, 
the preference for using fermions in this respect is quite reasonable, 
but there is also a good physical motivation for doing so. In particular, 
local fluctuations of topological charge are of interest because they are 
believed to cause light quarks to generate the $\eta'$ 
mass~\cite{ins_tHooft,WittenUA(1)} and the chiral condensate~\cite{ins_mixing}, 
for example. It is thus physically very natural to look for 
the imprint of these fluctuations in the low-lying Dirac eigenmodes 
which dominate the propagation of light quarks.

The concrete proposal of Ref.~\cite{Hor01A} is to investigate the behavior
of local chirality in the low-lying modes. This approach uses the fact that 
the instanton-like gauge fluctuations would leave a specific imprint in the 
individual modes, thus offering an opportunity to examine the consistency
of the instanton picture. More precisely, it was argued that if there are 
extended regions of (anti)self-duality in the gauge background, then the
probability distribution of the chiral orientation parameter 
$X$ ($X$-distribution) over the regions of strong field should exhibit 
double-peaked behavior in the vicinity of extremal chiralities. The initial 
study of the $X$-distribution in Ref.~\cite{Hor01A} indicated a flat behavior 
and thus a {\it qualitative} discrepancy with the instanton picture. However, 
later studies on finer lattices and/or with chirally symmetric fermionic 
actions~\cite{Followup,Gattr01A} revealed a visible degree of double-peaking.
This situation precludes any definite conclusions on purely qualitative grounds 
since the double-peaked structure is {\it necessary but not sufficient\/} for instanton 
dominance. It is a main purpose of this work to offer a more {\it quantitative} 
point of view.

In the first part of the paper (Sec.~\ref{csvsf}) we explain in detail why 
we choose to rely on the information encoded in low-lying Dirac modes, 
rather than using the conventional approaches for probing the gauge field 
topology on the lattice. We then concentrate on clarifying the possible
implications
of the $X$-distribution and argue that a double-peaked structure is not uniquely 
associated with self-duality. For example, if the gauge fields form approximately 
quantized~\footnote{When we refer to local quantization of topological 
charge we always imply quantization in units of $\pm 1$.} isolated
lumps of topological charge, then strong double-peaking is expected as well, 
even when such lumps are not self-dual. Also, some chiral peaking in the low
eigenmodes may occur even for non-quantized topological charge excitations
through some as-yet-unknown mechanism. The observed behavior of the 
$X$-distribution is thus not necessarily associated with an instanton-dominated 
vacuum. We suggest that the tentative conclusions of such nature be supplemented 
by quantitative comparison of QCD $X$-distributions to those for the existing 
ILM ensembles. The material in Sec.~\ref{csvsf} extends and complements the general 
discussion of Ref.~\cite{Hor01A}.

In the second part of this paper (Sec.~\ref{overlap_modes}), we present 
new results for $X$-distributions from eigenmodes of the overlap operator 
in Wilson gauge backgrounds at lattice spacings ranging from 
$a\approx 0.04\,\mbox{\rm fm}$ to $a\approx 0.12\,\mbox{\rm fm}$. 
In an attempt to verify whether the observed behavior can be ascribed 
to gauge structures with the parameters of the ILM, we determine the size 
and density of fermionic structures that contribute to the peaks of the  
$X$-distribution. This is motivated by the fact that for instanton-like 
excitations there is a direct correspondence between gauge and fermionic 
structure. Our consistency check leads to significant disagreement with the ILM. 
In fact, the lattice spacing dependence of the average radius (defined 
without reference to the t'Hooft zero mode profile which does not fit reasonably) 
indicates a finite continuum limit with value 
$R\approx 0.15\,\mbox{\rm fm}$, while the density of structures at 
$a\approx 0.04\,\mbox{\rm fm}$ is $n\approx 50\, \mbox{\rm fm}^{-4}$,
and we cannot exclude the possibility that it diverges in the continuum limit. 
We then point out that this not only disagrees with the ILM, but also suggests 
that the bulk of topological charge in QCD is not locally quantized in approximately
integer units. We give several arguments supporting this interpretation. 
Needless to say, if confirmed, this would have profound implications for 
the possible microscopic explanation of spontaneous chiral symmetry 
breaking. Finally, we give several arguments that our data is not contaminated 
by lattice artifacts usually referred to as ``dislocations'', and support
this conclusion by presenting the data for the Iwasaki gauge action. A preliminary 
version of this work is presented in compact form in Ref.~\cite{proceedings}.
 
\section{Cooling or Smoothing vs Fermions}
\label{csvsf}

The structure of equilibrium lattice Monte Carlo configurations provides 
a unique window for examining the nature of gauge fluctuations in the QCD 
vacuum. It also represents a distinctive way of thinking about how low-energy 
phenomena in QCD arise in terms of fundamental degrees of freedom. For example, 
as we will examine in some detail below, in the ILM mechanism 
of spontaneous chiral symmetry breaking the condensate arises due to mixing 
of t'Hooft's ``would be'' zeromodes associated with (anti)self-dual lumps, 
carrying approximately quantized topological charge. This line of reasoning
inherently assumes that this happens at the ``configuration level'' with the
vacuum dynamically generating the gauge potentials with these properties. 
If this is the case then the picture has a fundamental microscopic meaning.
Otherwise, it represents only a phenomenological modeling.

Unfortunately, examining lattice gauge fields directly leads to ambiguous
results, mainly due to the fact that fields are typically rough even
at the scale of a single lattice spacing. While the situation will improve 
with the approach to the continuum limit, this is probably not so when comparing 
the behavior at fixed physical distance. It is thus perhaps inevitable 
that some sort of filtering procedure be used to interpret the local 
structure of lattice gauge fields. In this section, we examine various
ways of approaching this issue.

Before proceeding to discuss this in more detail, it is useful to fix 
the language that will be used in what follows. In particular, we will
use the term {\it ``isolated lump''} or simply {\it ``lump''} to denote 
a local structure in the gauge field that can be enclosed by a hypersurface 
on which the field is approximately pure gauge. Consequently, the lump 
contains approximately integer-valued topological charge. A tendency 
towards local ``lumpy structure'' is one of the inherent properties 
of the ILM picture which distinguishes it from the situation when 
the gauge field fluctuates very inhomogeneously, forming peaks 
of topological charge density, but without any tendency for such quantization.

\subsection{Cooling}

One popular way of addressing the problem of short-distance fluctuations is 
the cooling method~\cite{cooling}. We will be rather generic when using this 
term, assuming just that it is a local minimization procedure for the gauge
action with initial state being the Monte-Carlo generated QCD configuration. 
We will not be concerned with various implementations or variations on 
the original idea.

With the elementary step being local, it is naturally expected that if the 
original configuration can be assigned a global topology in some way, then 
it will be preserved to a large extent during such a procedure. One can thus 
turn the argument around and to {\it define the global topological charge} 
of the original configuration through field-theoretic definition, let's say, 
on the corresponding cooled counterpart. Such a definition may not be entirely
satisfactory (the number of cooling sweeps is rather subjective and 
non-unique) but it can be made into a fairly well-motivated working scheme 
especially for the improved versions~\cite{imprcool}.

However, the situation is different if one is interested in the {\it local 
structure} of topological charge fluctuations. In the continuum, there are true 
local minima of the action for strictly self-dual or strictly anti-self-dual
fields. The superpositions of not overwhelmingly overlapping instantons and 
antiinstantons, while not exact minima, correspond to plateaus (or shallow 
valleys) in the action profile. After a few cooling sweeps when the gauge 
field undergoes large changes, the local minimization can bring the configuration 
to the vicinity of such a plateau. Here the fields become naturally smooth, 
the field-theoretic definition of topological charge leads to values that cluster 
around integers, and the evolution in the configuration space slows down. Further 
cooling possibly leads the configuration into another plateau, and eventually into 
the global minimum with nonperturbative fields completely removed~\footnote{For 
improved cooling it is in principle possible that the configuration remains 
indefinitely in the vicinity of local self-dual minimum (with 
instanton-antiinstanton pairs annihilated).}. 
By the nature of the argument, in the vicinity of the plateau the configuration 
will necessarily resemble a (multi)instanton-antiinstanton state regardless of 
the local properties of the original configuration. In other words, a possible 
observation of lumpy structure and local (anti)self-duality in the cooled 
configuration is to be expected, and cannot be used as an independent, logically 
satisfactory input for making conclusions about local fluctuations of topological 
charge or about the dynamical importance of instanton-like gauge fluctuations in
the QCD vacuum.

To illustrate this point, it is perhaps instructive to imagine that we are 
interested in studying the local structure of a sufficiently strongly-coupled 
lattice QCD vacuum, where instanton-like fluctuations are not expected to play 
any role. Consequently, an unbiased approach should find no traces of them. However, 
cooling the corresponding equilibrium configurations will still lead to plateaus 
and locally lumpy, self-dual behavior. Can we use this as a basis to conclude that 
instantons play a significant role in the strongly-coupled lattice QCD vacuum? 
Certainly not.
     
\subsection{Smoothing}

The inherent reason why cooling is a biased way of studying local properties 
of gauge fluctuations is that it uses the gauge action as a basis for the procedure.
This is not easily curable by improvements. However, one can also attempt to eliminate 
the unphysical short-distance fluctuations by some sort of {\it smoothing} without 
reference to the action. Cooling itself is a smoothing procedure in the sense that 
it replaces the original configuration with a smoother one. For the discussion in this 
section, however, we will be more specific in that respect, and call smoothing any 
procedure that can be interpreted as an action-independent space-time averaging of 
the fields. The prototype for this would be for example an APE-smearing~\cite{APE} 
that was actually used in this context~\cite{APEuse}.

While smoothing is a less invasive approach than cooling, the subjective element
in determining just how much smoothing is enough remains. This is troublesome 
because smoothing, by its nature, is also biased towards lumpy structure and 
can qualitatively change the local behavior. To see that, consider a very rough, 
inhomogeneous gauge field with various structures in it. Upon smoothing, the field 
varying over short distances will be quickly removed or reduced to a smooth 
four-dimensional ``bump'' if there is an underlying long distance structure 
present. Also, the lower-dimensional structures, such as ridges, sheets etc, will 
be eliminated soon by four-dimensional averaging. Thus, as the smoothing progresses, 
one naturally expects the stage when the configuration will be considerably lumpy 
with physical fields concentrated in the lumps. Further smoothing will cause 
the lumps to grow in size, to overlap, and eventually to reach a homogeneous 
situation with all physical fields removed. An unsettling question is where in 
this process one should stop and claim that the fields at the given stage represent 
the filtered local structure of the original configuration.

\subsection{Fermions}
\label{subsec:fermions}

Fortunately, there is a clean and meaningful strategy for approaching the above 
issues, namely to rely on the fermionic response to the corresponding gauge
background. To explain that, it is easiest to first think in terms of a theory 
in the continuum where a typical configuration is also expected to have a lot of 
ultraviolet fluctuations. Using fermions is quite plausible since we know that, 
for sufficiently smooth gauge fields, fermions reflect global topology 
exactly~\cite{AtiyahSinger} and for non-differentiable gauge fields the index 
of the Dirac operator can actually serve as an extended definition of topological 
charge. It is thus natural to expect that the local structure can be inferred 
from fermionic response as well, e.g. by studying the divergence 
of the flavor-singlet axial-vector current.

The new idea which makes the fermionic approach attractive is to look for the
imprints of topological charge fluctuations in the low-lying eigenmodes of the 
Dirac operator~\cite{Hor01A}. This is physically well-motivated, practical and, 
at the same time, has the potential to naturally solve the problem of ultraviolet
fluctuations without the subjective element involved. Indeed, the space-time 
structure of the low-lying modes is naturally smoother than that of the gauge 
fields themselves. This fact has two origins. First, the analogous Schr\"odinger-like 
eigenvalue problems typically yield smooth stationary states even when the underlying 
potential has discontinuous jumps. Secondly, among the eigenstates, the infrared 
modes are expected to be least sensitive to the short-distance features on top 
of the long-distance structure in the potential\footnote{Low-lying modes remain 
sensitive to the {\it isolated} small structures as they should.}. 
At the same time, it is necessary to realize that the underlying physics we
want to understand is the mechanism of how gauge fluctuations cause the light 
quarks to propagate in such a way as to form a quark condensate and to give the  
$\eta'$ its mass. In other words, the goal is to understand the dynamics underlying 
the propagation of light quarks. This physics is encoded in the low-lying Dirac 
modes~\cite{Hor01A}, and this is why we say that concentrating on these
modes (``fermion filtering'') is physically well motivated.

One possible approach is to infer the information about topological charge 
fluctuations from the individual low-lying modes~\cite{Hor01A}. This method 
is indirect and is motivated by the instanton picture of the vacuum. In other
words, the properties of instanton-like gauge fluctuations imply a specific 
qualitative behavior of the individual modes, thus giving an opportunity 
to check whether the observed structure is consistent with that scenario.
The specific proposal of Ref.~\cite{Hor01A} is to study the probability
distribution of the local chiral orientation parameter $X(n)$ defined by
\begin{equation}
    \tan\,\left(\frac{\pi}{4}(1+X) \right) = \frac{|\psi_L|}{|\psi_R|} \;,
    \label{eq:20}
\end{equation}
over the points $n$ where the eigenmode ($\psi_n^+\psi_n$) is large. Here 
$2\psi_L=(1-\gamma_5)\psi$, $2\psi_R=(1+\gamma_5)\psi$ are the left and right 
components of the eigenmode $\psi$. $X(n)$ is a local angle in the 
$|\psi_L|$-$|\psi_R|$ plane rescaled so that $X(n)=-1$ for purely right-handed
and $X(n)=+1$ for purely left-handed spinor $\psi_n$. For $\psi_L$,$\psi_R$
generated in a random independent fashion, the local orientation parameter
would be uniformly distributed between $-1$ and $+1$.

\subsection{The Implications of the $X$-distribution}

The usefulness of the $X$-distribution was discussed in detail in 
Ref.~\cite{Hor01A}, where it was argued that if the gauge background contains 
extended regions of strong (anti)self-dual fields, then the $X$-distribution should 
exhibit peaks near the extremal values. This is a consequence of the fact that 
in the eigenvalue problem for the Dirac operator, the (anti)self-dual part of the
gauge field enters as a potential term for the (right)left component of the
eigenmode. 

Here we would like to emphasize that this is not a unique way for the double-peaked 
behavior to arise. To illustrate that, consider subjecting the fermion to a background 
configuration that is lumpy in the sense that we have defined it here,
i.e. consisting of relatively isolated lumps surrounded by approximately field-free 
regions. (Such configurations can be artificially prepared irrespective of whether 
they resemble typical equilibrium configurations of QCD.) The fields comprising 
the lumps have no definite duality properties but yet, carry approximately quantized 
topological charge. A particular lump can thus be thought of as an individual object 
and, in the absence of all the other lumps, it would induce zero modes of the Dirac 
operator with chirality dictated by the index theorem. For simplicity, consider 
a situation with just two lumps with topological charges $Q=\pm 1$. The residual 
interaction will cause these ``would-be'' zero modes to mix and, to a first 
approximation, the true eigenmodes would be the two linear combinations of the strictly 
chiral ``would-be'' zeromodes localized on the lumps. In the resulting {\it topological}
near-zero modes, the $X$-distribution will be strongly peaked at $\pm 1$.

What we have described above is the same scenario that is a basis for the ILM 
mechanism of spontaneous chiral symmetry breaking~\cite{ins_mixing}. The point is 
that self-duality is not essential for this argument since we didn't need to invoke 
instantons at all. It is only the local quantization of topological charge that 
matters and yet, the peaked $X$-distribution is expected, as well as spontaneous 
chiral symmetry breaking if such configurations turned out to be dynamically important.

We emphasize this point to show that the qualitative double-peaked behavior of the 
$X$-distribution alone does not in itself provide a basis for the verification of the 
instanton picture. It is a necessary but certainly not a sufficient condition for 
such a conclusion. Apart from the above argument, there can possibly be other 
mechanisms for producing the peaked distribution (as well as spontaneous chiral 
symmetry breaking). In fact, we will argue later that there are reasons to believe 
that topological charge is actually not locally quantized in QCD. Nevertheless, 
the study of {\it quantitative} characteristics of the $X$-distribution can still be 
very useful for distinguishing various scenarios as well as for other 
purposes~\cite{Gattr01A}.

\subsection{Lattice Fermions}

When considering the lattice-regularized theory we have the usual extra 
freedom in choosing the lattice action. According to the standard universality
assumption, all local actions with appropriate symmetries and correct classical 
behavior should give consistent results sufficiently close to the continuum
limit. However, comparing results from different discretizations at finite 
lattice spacing usually requires some care. When asking questions about 
the nature of fluctuations in the pure gauge vacuum, it is natural to fix a 
particular lattice gauge action and study the theory at different lattice 
spacings to be able to extrapolate to the continuum limit. Moreover, if we choose 
to examine the nature of these fluctuations using a lattice fermion, then it is 
essential that the fermionic action be kept fixed as well, even though it only 
plays the role of a probe.

To illustrate this point, consider some lattice Dirac operator $D(U)$ 
and the set of related operators $D_n(U) \equiv D(U_n)$, where $U_n$ is the gauge 
configuration obtained from $U$ by performing $n$ smearing steps (e.g. APE steps)
with other parameters of the smearing procedure fixed~\footnote{See 
Ref.~\cite{deGrand00A} for example.}. If $D(U)$ is an acceptable lattice Dirac 
operator, then $D_n(U)$ is acceptable as well, assuming that $U_n$ exists. However, 
if we investigate the local properties in the equilibrium ensemble $\{ U \}_{eq}$ 
using $D_n$, then the answers we get will clearly depend on $n$. In particular, 
$D_{1000}$ will probably indicate much larger structures in the gauge field than 
$D_1$. Nevertheless, one is in principle allowed to use $D_{1000}$ 
as long as it is used consistently in {\it meaningful} extrapolations 
to the continuum limit, which might be difficult. However, one can hardly conclude 
anything meaningful about the behavior in the continuum limit by comparing the 
results from $D_1$ at one lattice spacing and from $D_{1000}$ at a different 
lattice spacing. 
 
Regarding the choice of the fermionic action, we would like to point out one 
particular aspect illustrated by comparing the two extremes represented by the 
Wilson-Dirac operator and the overlap operator based on it. The overlap operator 
has exact lattice chiral symmetry at finite lattice spacing, allowing for 
continuum-like theoretical analysis with respect to chiral symmetry and topology. 
On the other hand, it is non-ultralocal and, due to extended gauge connections, 
effectively samples the gauge field potential over nonzero physical 
distance~\footnote{Due to this ``chiral smoothing'' the features smaller than some 
physical threshold will not be resolved at finite lattice spacing. The results 
of Ref.~\cite{Gattr01B} are a manifestation of this fact.}. 
This should not matter for the continuum limit, based on the expectation that the 
operator is local over equilibrium ensembles of Wilson pure gauge theory 
at sufficiently weak coupling. The Wilson-Dirac operator, on the other hand, 
has mutilated chiral properties and when using it in this context, one relies 
heavily on the assumption that this problem will go away in the continuum limit. 
At the same time, the Wilson-Dirac operator has a perfect resolution in the sense 
that there is no additional smoothness beyond the fact that we end up inspecting 
the infrared eigenmodes. It would be very interesting to know whether the two 
actions provide for consistent continuum extrapolations for quantities related 
to topology. In this work, we use an overlap operator which, even though 
expensive to implement, is very convenient for its theoretical advantages.

The above considerations also suggest that there may be a more general complementarity 
between the degree of chiral symmetry and the resolution for lattice fermionic actions. 
Indeed, if we start with a maximally ultralocal operator, we can try to improve its 
chiral properties by including couplings at larger distances and adding more 
complicated gauge connections, but by doing so we worsen the resolution. On the other 
hand, if one starts with the operator with exact lattice chiral symmetry, then this 
operator must be non-ultralocal~\cite{nonultr}, and an attempt to improve its 
resolution by dropping couplings at large distances will result in deteriorated 
chiral properties.

\section{Low-Lying Modes of the Overlap Operator}
\label{overlap_modes}

We have calculated and examined the low-lying modes of the overlap operator
in Wilson gauge backgrounds over a wide range of lattice spacings. The massless 
overlap operator~\cite{Neu98BA}
\begin{equation}
D \;=\; \rho\, \Biggl[\; 1 \,+\, (D_W-\rho) \, 
              {1 \over \sqrt{(D_W-\rho)^+(D_W-\rho)}}\;\Biggr]  
\end{equation}
with $D_W$ the Wilson-Dirac operator and $\rho=1.368$ ($\kappa=0.19$) is used 
throughout this paper. The rational approximation for the matrix sign 
function~\cite{rational,kyoverlap} was utilized to implement $D$ and a small number 
of eigenmodes of $\gamma_5 D_W$ were projected out to both increase 
the accuracy of the approximation and to speed up the convergence. 

We used the Ritz variational method~\cite{Ritz} to obtain the low-lying 
eigenmodes of $H^2\equiv D^+D$. As a consequence of normality (\,$[D,D^+]=0$\,) and 
$\gamma_5$-hermiticity (\,$D^+=\gamma_5 D \gamma_5$\,), the nonzero low-lying modes 
of $H^2$ are doubly degenerate and the eigenmodes of $D$ can be constructed in the
corresponding subspaces since $[H^2,D]=0$. Also, $H^2$ is proportional to the chirally 
non-symmetric part of $D$, implying that $[\gamma_5,H^2]=0$. This allows for 
diagonalization in separate chiral sectors in different runs, thus speeding up the 
process and easing the memory requirements for large lattices. The typical accuracy 
of calculated eigenvalues (as measured by differences of complex-conjugate pairs) 
is about one part in $10^5$ or better.

To avoid confusion, it is useful to fix the language that will be used when 
discussing the local properties of a given eigenmode $\psi$ of $D$. We will 
refer to
\begin{equation} 
   d(n) \,\equiv\, \psi_n^+\psi_n                \qquad\qquad\qquad
   c(n) \,\equiv\, \psi_n^+ \gamma_5 \psi_n
\end{equation} 
as {\it density} and {\it chirality} respectively, and to $X(n)$ of 
Eq.~(\ref{eq:20}) as {\it chiral angle}. Since $d(n)$, $c(n)$ and $X(n)$ are identical 
for eigenmodes corresponding to complex conjugate
nonzero eigenvalues $(\, \psi_{\lambda^*} = \gamma_5\psi_{\lambda})$, we will
treat $\psi_{\lambda}$, $\psi_{\lambda^*}$ as a pair.    

The parameters of the Wilson gauge ensembles used are listed in 
Table~\ref{ensemb_tab}. Configurations for the three finest lattices are separated by 
twenty thousand sweeps. The quoted values of lattice spacings were obtained 
from the Sommer parameter using the interpolation formula given in 
Ref.~\cite{Sommer}. Our finest lattice spacing is outside the interpolation range 
and we linearly extrapolate from $\beta=6.5$. Note that the physical volumes involved 
are such as to contain on average 3-4 (anti)instantons if the ILM scenario is relevant,
ensuring that the mixing of 't Hooft zero modes would take place. For all 
the configurations we have calculated the eigenfunctions for the zeromodes and 
at least two pairs of near-zeromodes. 

\begin{table}[b]
  \centering
  \begin{tabular}{cccc}
  \hline\hline
  \multicolumn{1}{c}{$\beta$}  &
  \multicolumn{1}{c}{$a$ [fm]}  &
  \multicolumn{1}{c}{$V$}  &
  \multicolumn{1}{c}{\# configs}  \\
  \hline
  5.85  & 0.123 & $10^3$x$20$ & 12 \\
  6.00  & 0.093 & $14^4$      & 12 \\
  6.20  & 0.068 & $20^4$      & 8  \\
  6.55  & 0.042 & $32^4$      & 5 \\
\hline \hline
\end{tabular}
\caption{Ensembles of Wilson gauge configurations.}
\label{ensemb_tab} 
\end{table}

\subsection{Results for the $X$-distribution}

\begin{figure}
\begin{center}
\centerline{
\epsfxsize=8.0truecm\epsffile{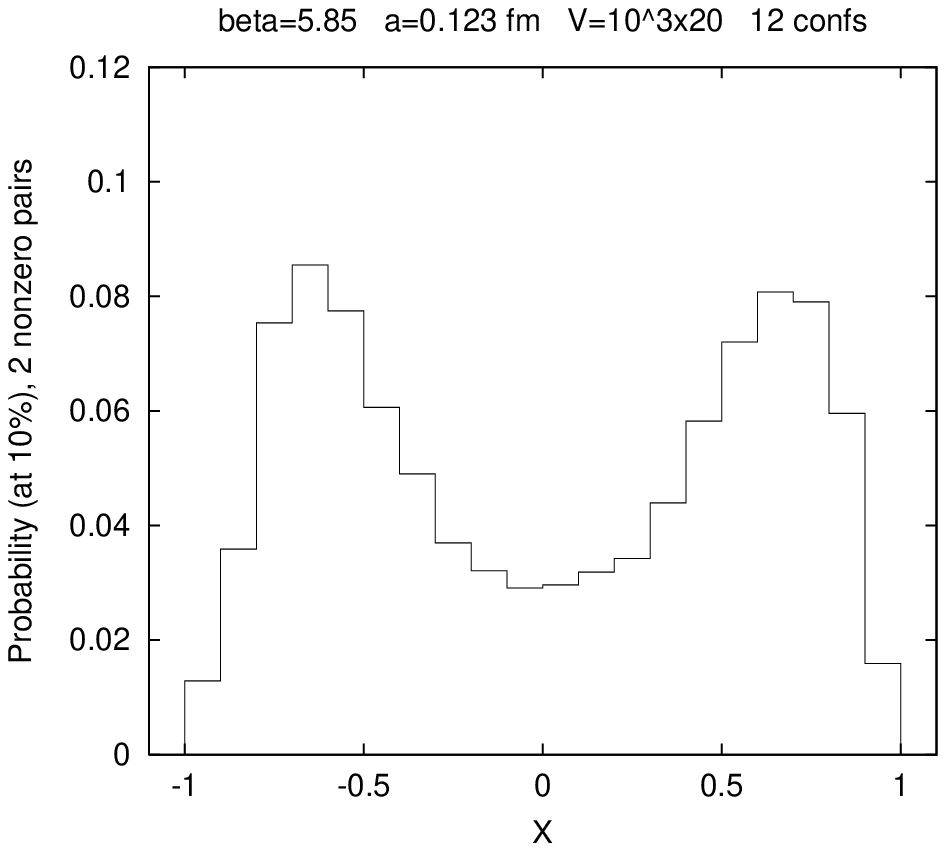}
\epsfxsize=8.0truecm\epsffile{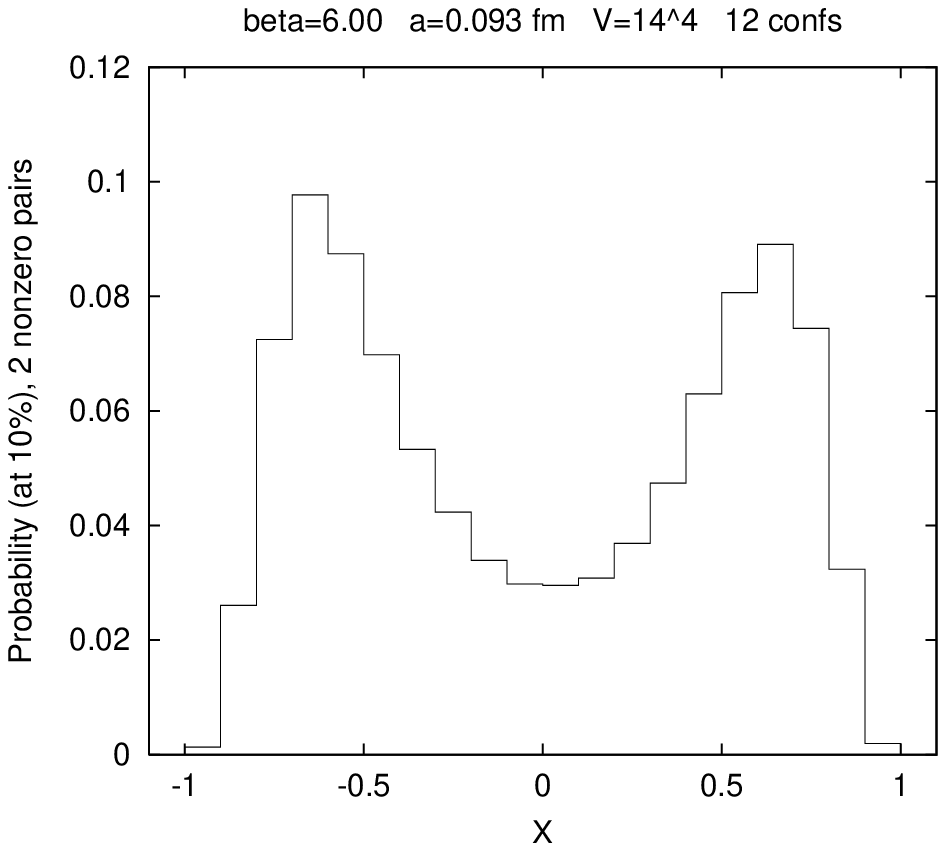}
}
\vskip 0.15in
\centerline{
\epsfxsize=8.0truecm\epsffile{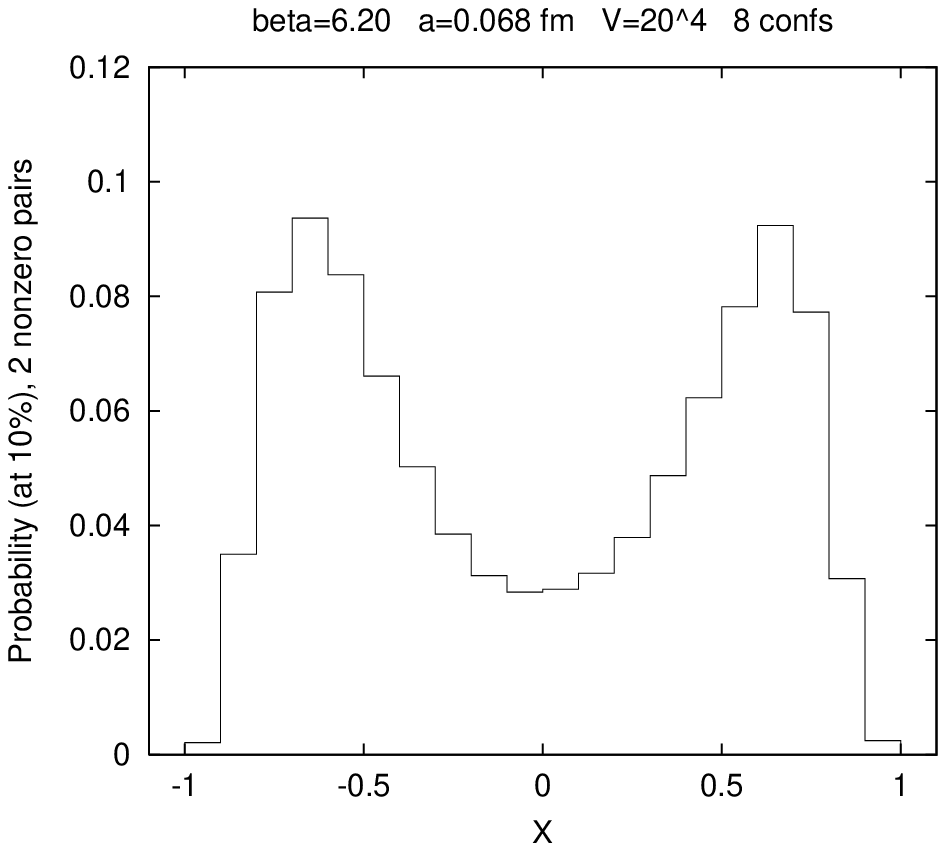}
\epsfxsize=8.0truecm\epsffile{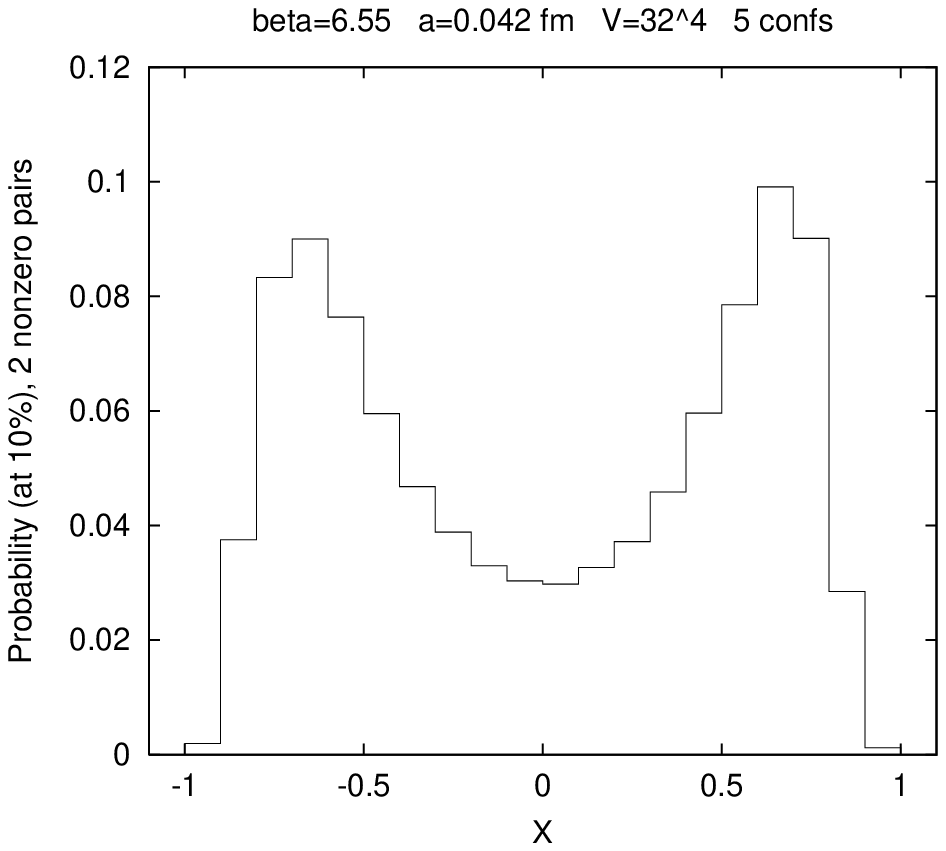}
}
\caption{$X$-distribution for four Wilson gauge ensembles considered.}
\label{Xdists}
\end{center}
\end{figure}

To make meaningful comparisons for the behavior of the $X$-distribution at 
different lattice spacings and different lattice sizes, it is necessary to 
fix the fraction $f$ of the points that are examined on each lattice. In 
fact, it would be more appropriate to always speak of $X_f$-distributions.
For a given low-lying mode we order the lattice sites by the magnitude of 
density and consider the top $f V$ points, where $V$ is the lattice volume.
It is assumed that the underlying gauge field is strongest in the regions
so selected~\cite{Hor01A}. For most of the results discussed here we fixed
$f=0.1$. This is mostly motivated by the fact that we intend to relate our 
findings to the ILM. Diakonov and Petrov~\cite{ILM} use the packing fraction 
$\rational18$ in their theoretical arguments. On the other hand, if one 
naively calculates the packing fraction as $f=n V_\rho$, where $n$ is the 
density in fm$^4$ and $V_\rho$ the volume of a four-dimensional sphere of 
radius $\rho$, then one obtains approximately $1\over{20}$ if ILM values 
are used. We thus view $f={1\over{10}}$ as a reasonable compromise. The 
qualitative and even quantitative conclusions that we will make do not depend 
on the precise value from the range quoted above as long as it is fixed.

\begin{figure}
\begin{center}
\centerline{
\epsfxsize=0.7\hsize\epsffile{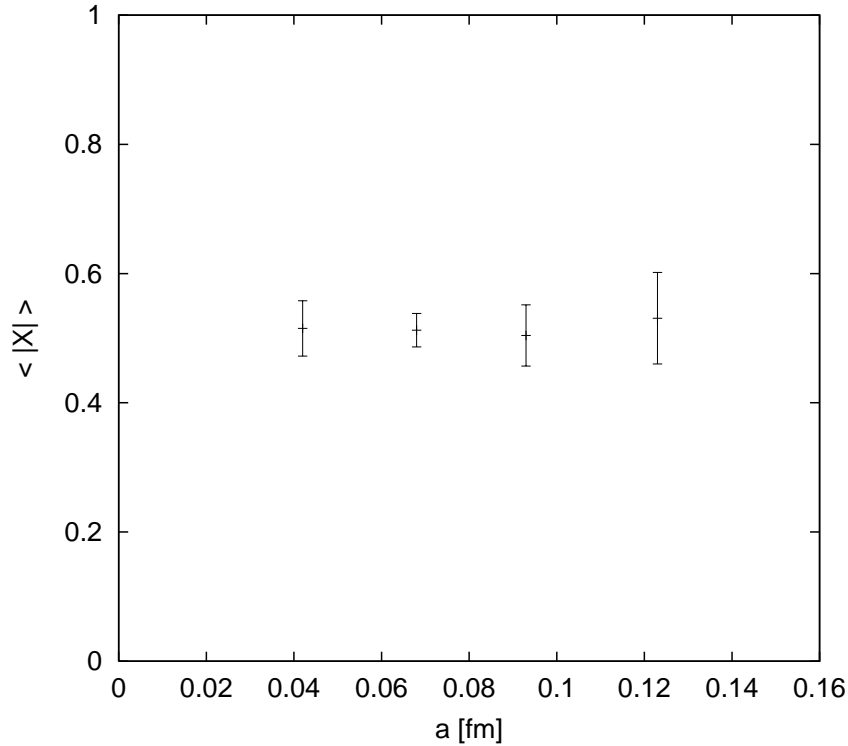}
}
\caption{Dependence of $< |X| >$, Eq.~(\ref{eq:40}), on the lattice spacing. 
The quoted uncertainty represents a rough estimate using the asymmetry 
$|p_{X_i} - p_{-X_i}|$ as a basis for the calculation.} 
\label{absXs}
\end{center}
\end{figure}

Our results for the $X$-distribution from the lowest two nonzero pairs of modes 
included for each configuration are shown in Fig.~\ref{Xdists}. The histograms
are normalized so that the sum of the values in all bins adds up to unity.
There are visible peaked maxima at $X\approx \pm 0.65$, but there appears 
to be very little change across the wide range of lattice spacings studied. 
To quantify this observation, we have calculated the average value of 
$|X|$ with the histograms serving as a probability profile, i.e. 
\begin{equation}
< |X| > \; \equiv \, \sum_i p_{X_i}\, |X_i|
\label{eq:40}
\end{equation}
where $p_{X_i}$ is the value at the bin 
with center $X_i$. For truly peaked distribution, this should reflect the 
approximate position of the peak. The results are shown in Fig.~\ref{absXs} 
indicating a very flat behavior as a function of the lattice spacing. 
The underlying dynamics thus does not appear to generate more chiral peaking 
closer to the continuum limit, nor do the positions of the peaks move 
appreciably closer to $\pm 1$.

\begin{figure}
\begin{center}
\centerline{
\epsfxsize=0.7\hsize\epsffile{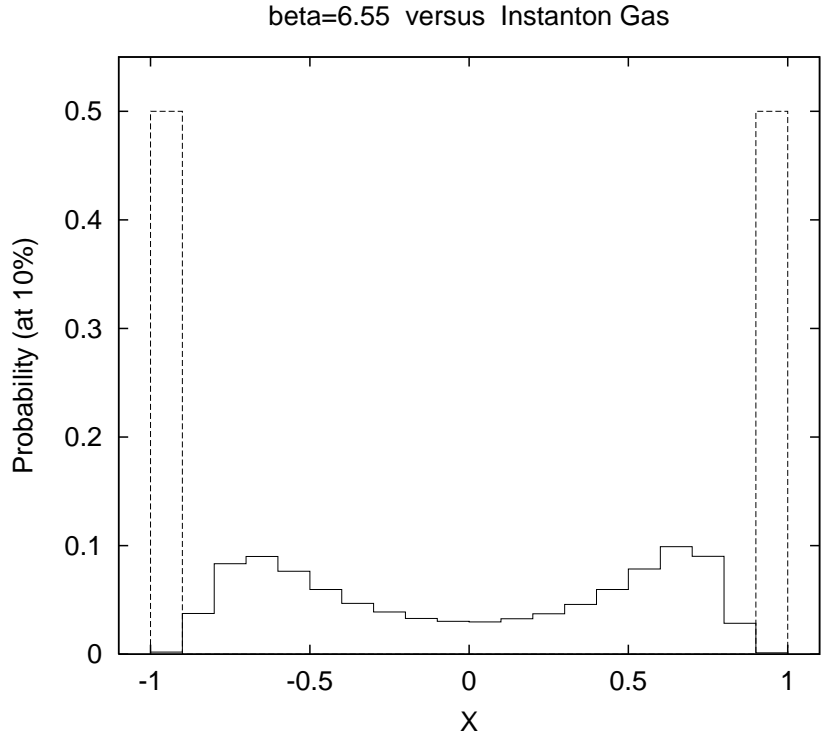}
}
\caption{Comparison of $X$-distributions at $\beta=6.55$ for near-zero modes 
(solid line) and exact zero modes (dashed line) using the overlap Dirac 
operator.} 
\label{b655bench}
\end{center}
\end{figure}

As we have emphasized in Sec.~2, the qualitative observation that the
$X$-distribution exhibits peaked maxima at nonzero $X$ is not sufficient for 
concluding that the instanton picture of topological charge fluctuations in 
the QCD vacuum is correct. However, the study of quantitative characteristics 
of the distribution might still be very illuminating. We illustrate this by 
comparing in Fig.~\ref{b655bench} the $X$-distributions for the near-zero 
modes and for the exact zeromodes. The latter is the same distribution that 
would be expected in the limit of a very dilute instanton gas (or a dilute 
gas of arbitrary topologically nontrivial lumps). It is clear that, even 
at our smallest lattice spacing, the $X$-distribution of the near-zero modes 
is far less chirally peaked than that of the exact zero modes\footnote{This 
is equally well reflected in the fact that 
we found typical values $< |X| > \approx 0.52$ (see Fig.~\ref{absXs}), while 
at the same time $< |X| >=1$ for distribution strictly peaked at $X=\pm 1$.},
confirming the prevalent view that the dilute instanton gas picture is not 
realistic~\cite{ILM}.
This also suggests that one reliable way to verify whether lattice QCD 
$X$-distributions are consistent with the ILM prediction is a direct 
comparison to distributions from existing ILM ensembles.

\subsection{Structures in the Gauge Field}

Our next goal is to verify whether the observed $X$-distribution can be
attributed to the underlying local structures with properties similar 
to those assumed in the ILM. If the low-lying mode arises from mixing 
of t'Hooft ``would be'' zero modes associated with ILM instantons, then 
this mode inherits the lumpy structure of the underlying gauge field, 
with lumps being of a prescribed size, shape and abundance.

\begin{figure}
\begin{center}
\centerline{
\epsfxsize=0.7\hsize\epsffile{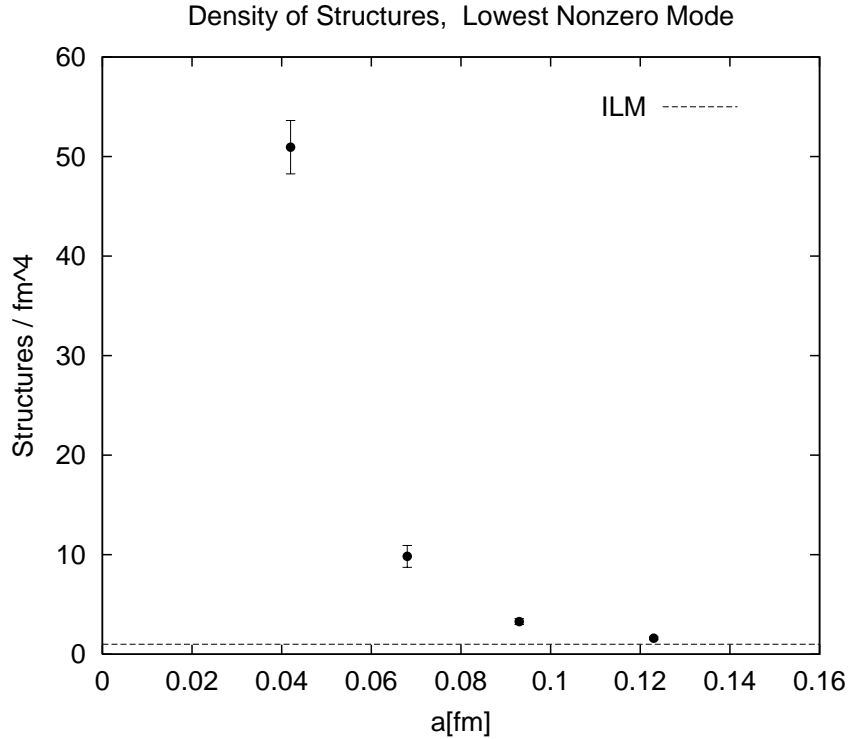}
}
\caption{Density of structures (in fm$^{-4}$) as a function of the lattice 
spacing.} 
\label{density}
\end{center}
\end{figure}

We identify such {\it possible} structures by finding the local maxima of 
density, $d(n)$, in the mode, and requiring that the profile of density around 
them resembles a four-dimensional peak at least in some average sense. To be 
more precise, we start by locating the maxima over the distance $\sqrt{3}$, 
i.e. finding the set 
\begin{equation}
   {\cal M} \equiv \{\, n\,:\, d(n) > d(m), \,|n-m|\le \sqrt{3}\,\}
\end{equation} 
For these candidates, specified by their position $n$, we then compute the 
functions
\begin{equation}
    d_{n,\mu}(r) \,\equiv\; < \,d(m)\, >_{{|n-m|=r}\atop {(n-m).\muhat>0}}
    \qquad\quad
    \mu=\pm 1,\ldots,\pm 4,\; r>0
    \label{eq:80}
\end{equation}
representing the average of $d(m)$ over the spherical shell of radius $r$ centered 
at $n$, restricted to the points for which $(n-m)$ has a component in the 
$\mu$-direction. Local maximum is retained only if $d_{n,\mu}(r)$ decays 
monotonically from origin over the distance $\sqrt{3}$ for all directions $\mu$. 

Note that the motivation for choosing $\sqrt{3}$ as a reference lattice 
distance in the above procedure is that it is a minimal distance for which
the accidental occurrence of a structure can be statistically excluded for
the largest lattice volumes we are working with. In other words, if we were 
to generate the density on sites of a $32^4$ lattice using random numbers, 
no structures would typically be found using the above prescription. 
Conversely, if a structure is identified in a low-lying mode, we take it as 
a signal that a nontrivial fluctuation of the gauge field is causing its 
occurrence.

\begin{figure}
\begin{center}
\centerline{
\epsfxsize=5.4truecm\epsffile{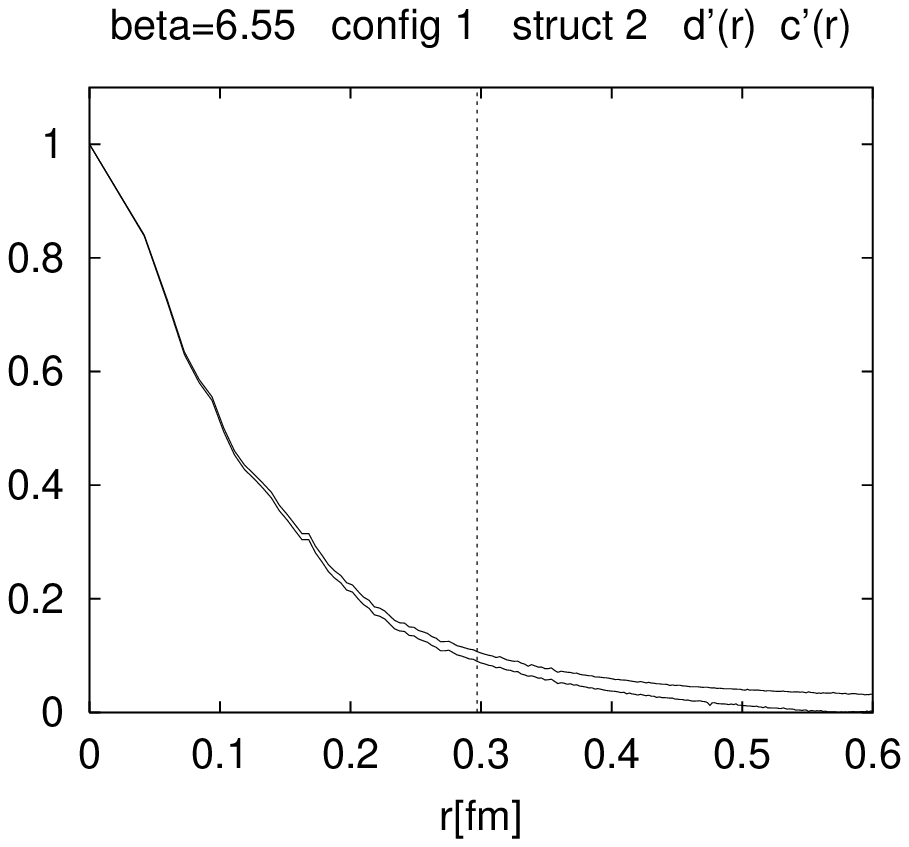}
\epsfxsize=5.4truecm\epsffile{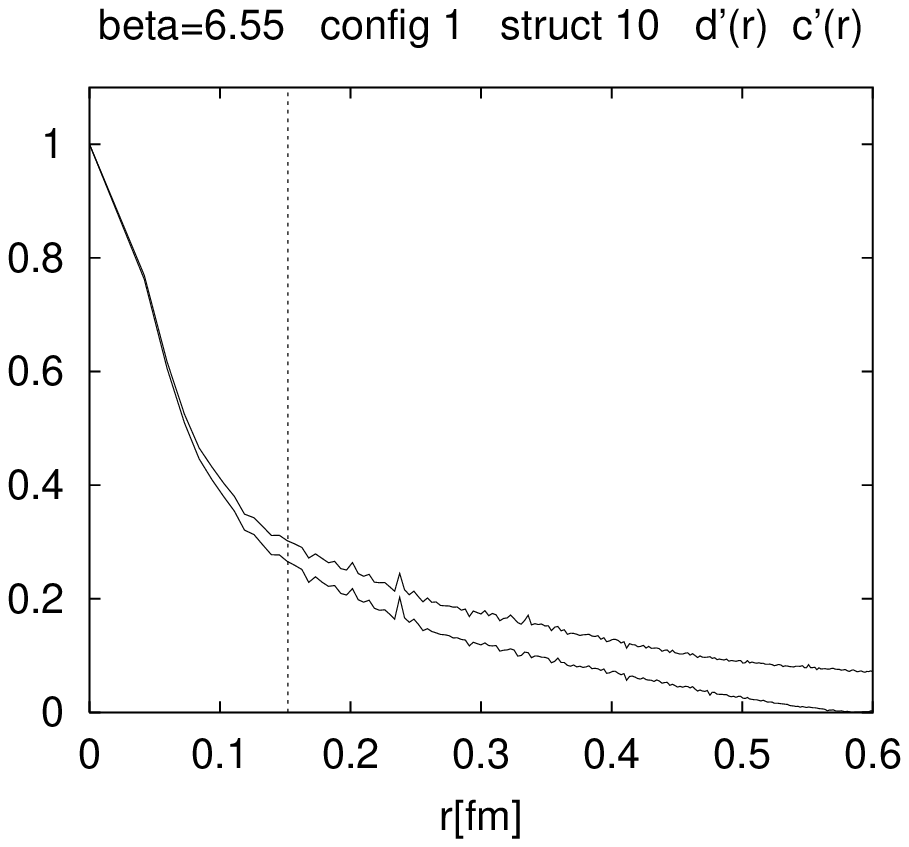}
\epsfxsize=5.4truecm\epsffile{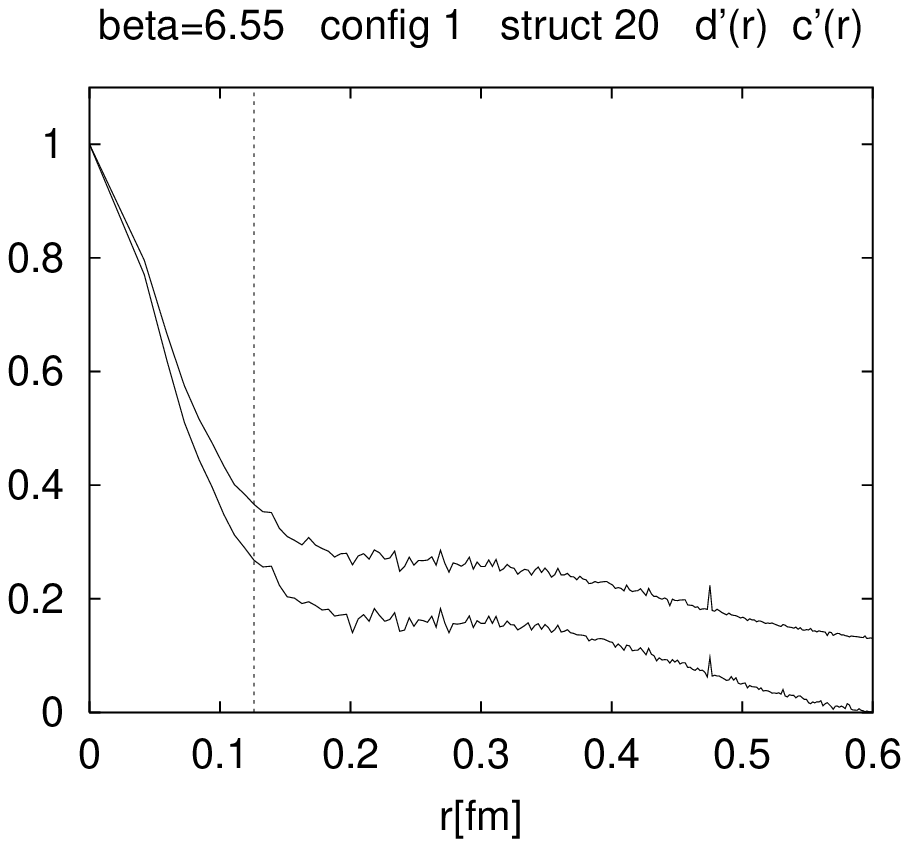}
}
\vskip 0.15in
\centerline{
\epsfxsize=5.4truecm\epsffile{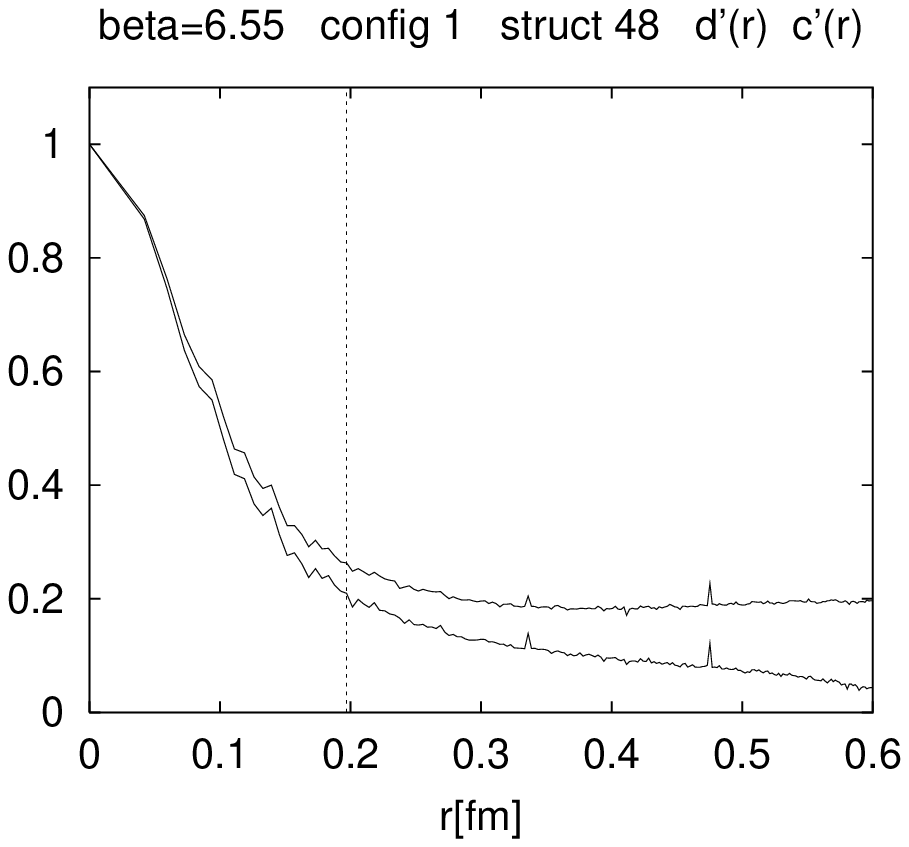}
\epsfxsize=5.4truecm\epsffile{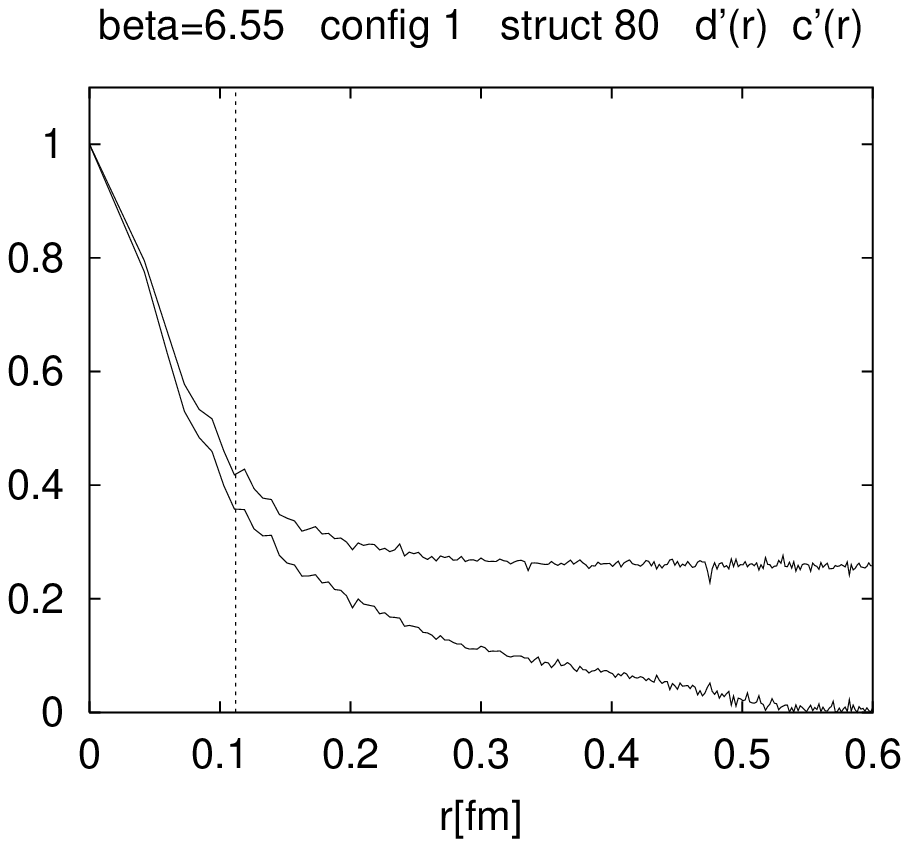}
\epsfxsize=5.4truecm\epsffile{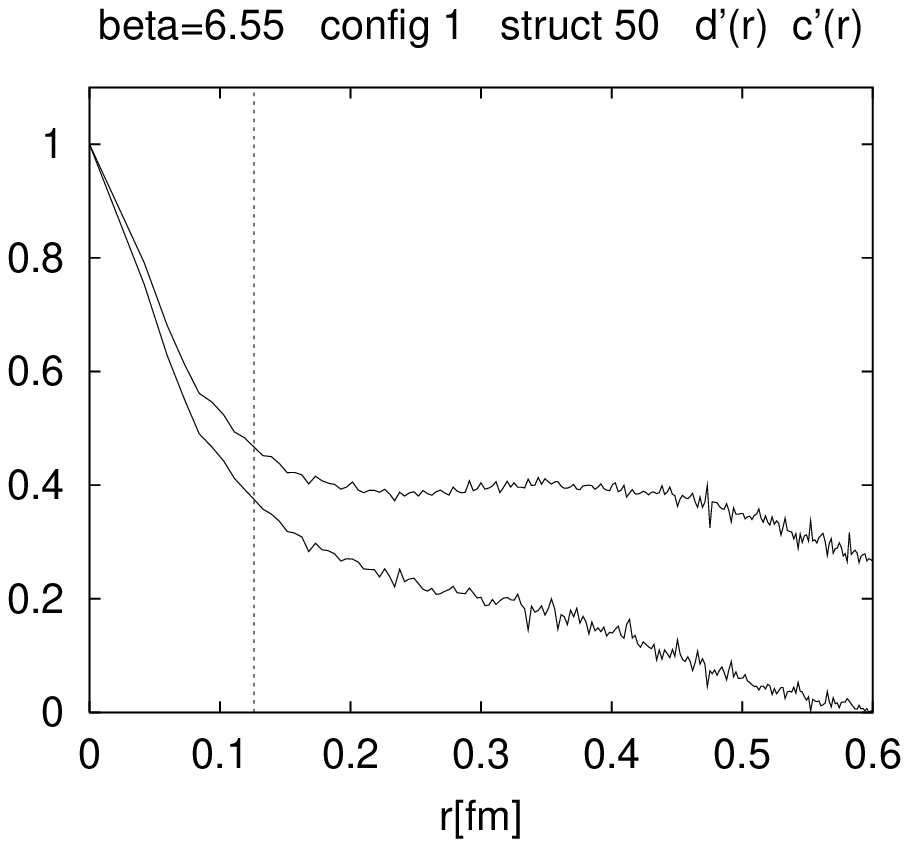}
}
\vskip 0.15in
\centerline{
\epsfxsize=5.4truecm\epsffile{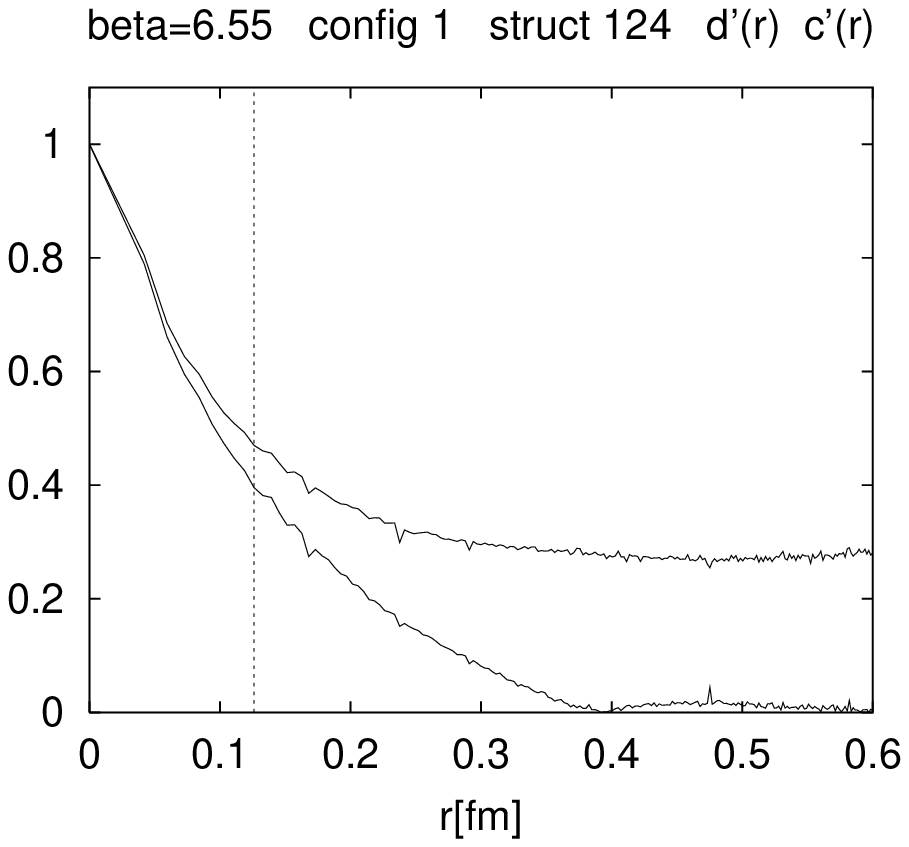}
\epsfxsize=5.4truecm\epsffile{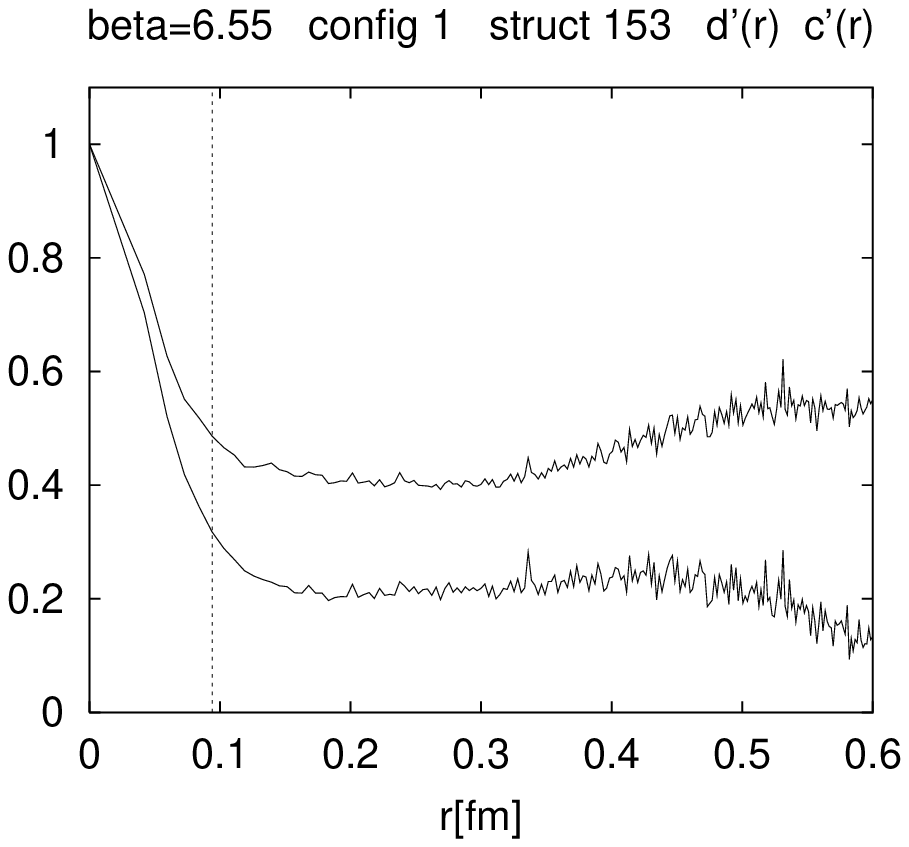}
\epsfxsize=5.4truecm\epsffile{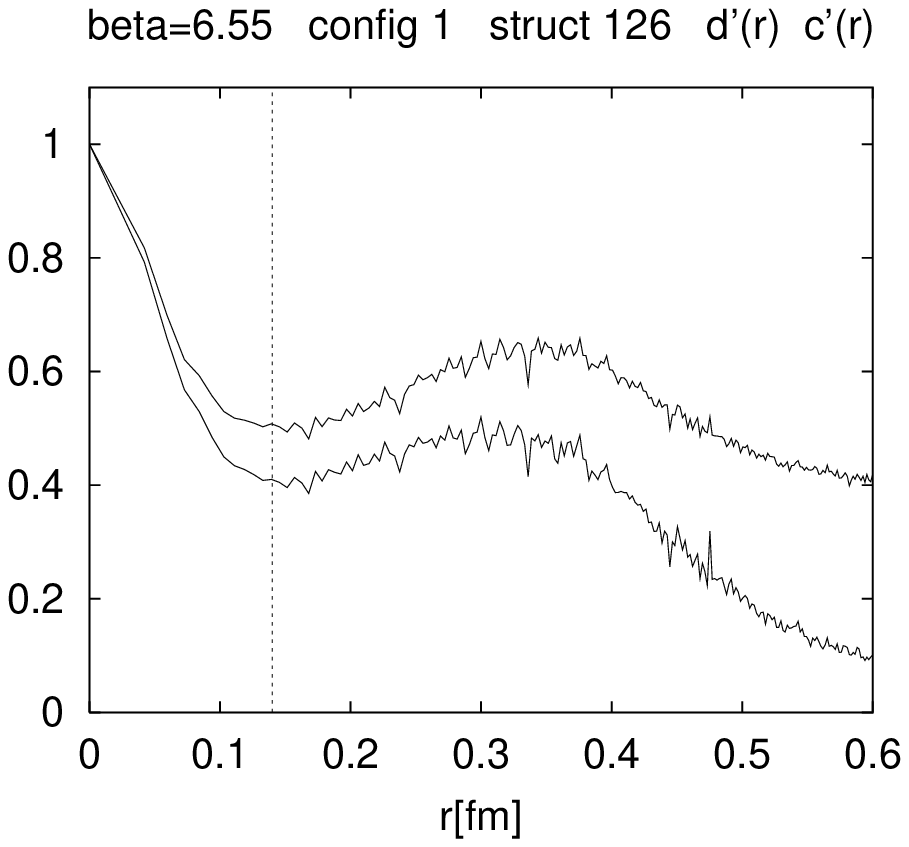}
}
\caption{The sample behavior of $d_n(r)/d_n(0)$ (upper curve) and $c_n(r)/c_n(0)$ 
(lower curve) for typical structures at $\beta=6.55$. The radius, $R_n$, of the 
corresponding region of coherent local chirality is shown as a vertical line. 
The figures are ordered so that the intensity of the peaks decreases vertically 
and the quality (degree of isolation) of peaks decreases horizontally.}
\label{Structures}
\end{center}
\end{figure}

Finally, from the structures identified in this way, only those were selected
that have a chance of contributing to the peaks of the $X$-distribution. 
To make this final cut, we only retain structures for which the corresponding 
center $n$ is among the 10\% of points with highest density ($f=0.1$ was used to
generate $X$-distributions), and for which the chiral angle satisfies
$|X(n)| \ge 0.5$ as suggested by the position of the maxima in Fig.~\ref{Xdists}.
The resulting density of structures as a function of lattice
spacing is shown in Fig.~\ref{density}, indicating a large disagreement with 
the ILM assumption at smaller lattice spacings. We will return to the 
interpretation of this behavior later.

\subsection{Sizes from Coherent Regions of Local Chirality}

To study the typical sizes of the structures, we first use the definition 
that is motivated only by the assumed lumpy behavior of the gauge field rather 
than the specific profile of an instanton. The mixing of ``would be'' zero modes
in the lumpy background would produce regions of coherent local chirality
in the near-zero modes, concentrated around the maximum of the lump. We thus
define the radius of the structure at point $n$ as the radius of the largest
hypersphere centered at $n$, containing points with the same sign of local
chirality, namely
\begin{equation}
    R_n \,\equiv\, \max \, \{\, r \,:\, c(n)\, c(m) > 0,\, 
                                    |n-m|\le r \,\}
    \label{eq:100}
\end{equation}
To assess the feasibility of this definition, we compute functions
\begin{equation}
    d_n(r) \,\equiv\, < \,d(m)\, >_{|n-m|=r}
    \qquad\qquad
    c_n(r) \,\equiv\, < \,c(m)\, >_{|n-m|=r}
    \label{eq:120}
\end{equation}
representing the average density and chirality over the spherical shell of radius 
$r$ centered at $n$.
In Fig.~\ref{Structures} we show the sample behavior of functions $d_n(r)/d_n(0)$ 
and $c_n(r)/c_n(0)$ for typical structures, and how this relates to the radius 
assigned by our definition (vertical line in Fig.~\ref{Structures}). The examples 
of structures with high, medium and low intensity (density) are put in the top, 
middle and bottom row respectively. Note that the level of ``background'' increases 
for lower intensities. Also, the structures with cleaner peaked behavior 
(typically more isolated ones) are put on the left and quality is decreasing 
to the right. The examples on the right represent the ``worst'' cases we have
identified. From these examples it can be seen that the peaks of density and 
chirality are well contained within the determined radius. 

\begin{figure}
\begin{center}
\centerline{
\epsfxsize=0.7\hsize\epsffile{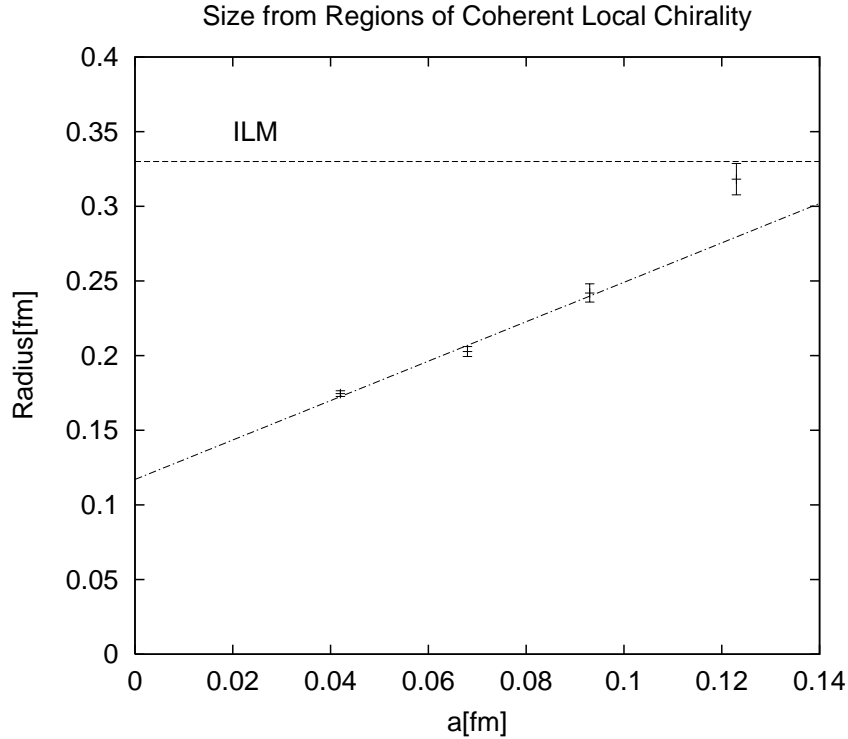}
}
\caption{Average radius, $< R_n >$, of structures from regions of coherent 
local chirality. The lowest nonzero mode was used for calculation. Data for the three 
smallest lattice spacings were used to obtain the fit. The horizontal line represents 
the radius of an ILM instanton.} 
\label{Sizechir}
\end{center}
\end{figure}

\begin{figure}
\begin{center}
\centerline{
\epsfxsize=0.7\hsize\epsffile{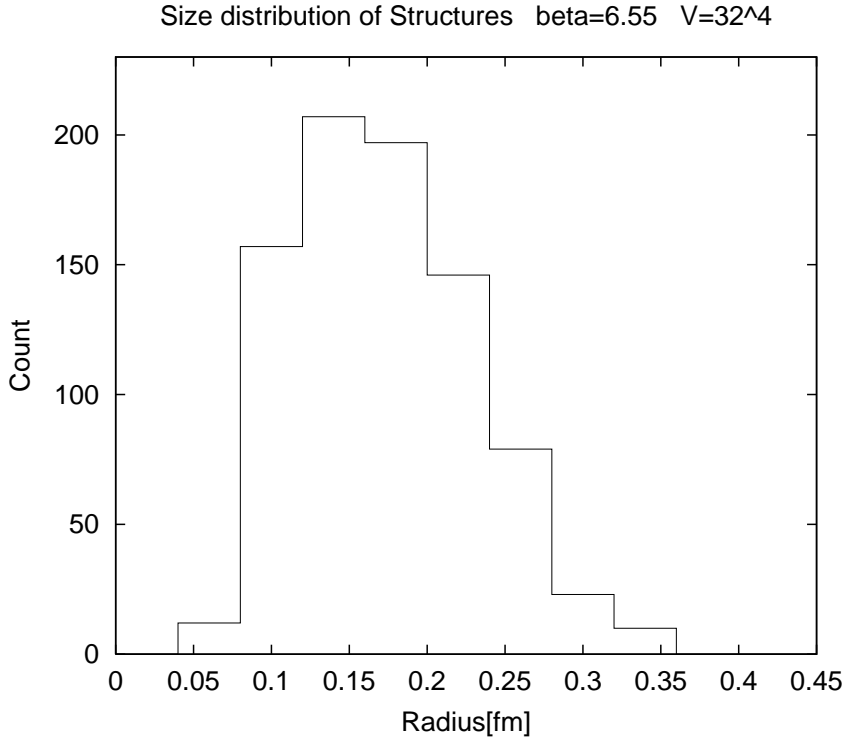}
}
\caption{The distribution of sizes $R_n$ at $\beta=6.55$.}
\label{Axsizdist}
\end{center}
\vskip -0.3in
\end{figure}

Using the definition (\ref{eq:100}), we have studied the average sizes of 
structures as a function of the lattice spacing with results shown in 
Fig.~\ref{Sizechir}. The straight line is a fit using the three smallest lattice 
spacings considered. The average radius at $a\approx 0.12\, {\mbox{\rm fm}}$ is close 
to the ILM value and consistent with a recent estimate at a similar lattice 
spacing~\cite{deGrand00A}. However, it would be erroneous to conclude agreement with
the ILM based on the data at single lattice spacing; the average value decreases 
significantly for finer lattices and the continuum-extrapolated estimate is in 
striking disagreement with the ILM. There appears to be a positive curvature in our data 
and a well-defined finite value (with rough estimate of about 
$0.15\, {\mbox{\rm fm}}$) in the continuum limit. The fact that our procedure leads 
to a finite size in physical units is significant because it characterizes the physical 
size of regions of coherent local chirality that are necessary to develop the large 
$\eta'$ hairpin correlator required to solve the $U(1)$ problem~\cite{Hor01A}. We  
also emphasize that while our definition of size was motivated by assuming 
the lumpy structure of topological charge fluctuations, it makes very good sense even 
if the structures do not carry the quantized values of topological charge, and are not 
related to instantons. The distribution of sizes at $\beta=6.55$ is shown in 
Fig.~\ref{Axsizdist}.

\subsection{Sizes from Instanton-Assumed Profiles of Density}

\begin{figure}
\begin{center}
\centerline{
\epsfxsize=8.0truecm\epsffile{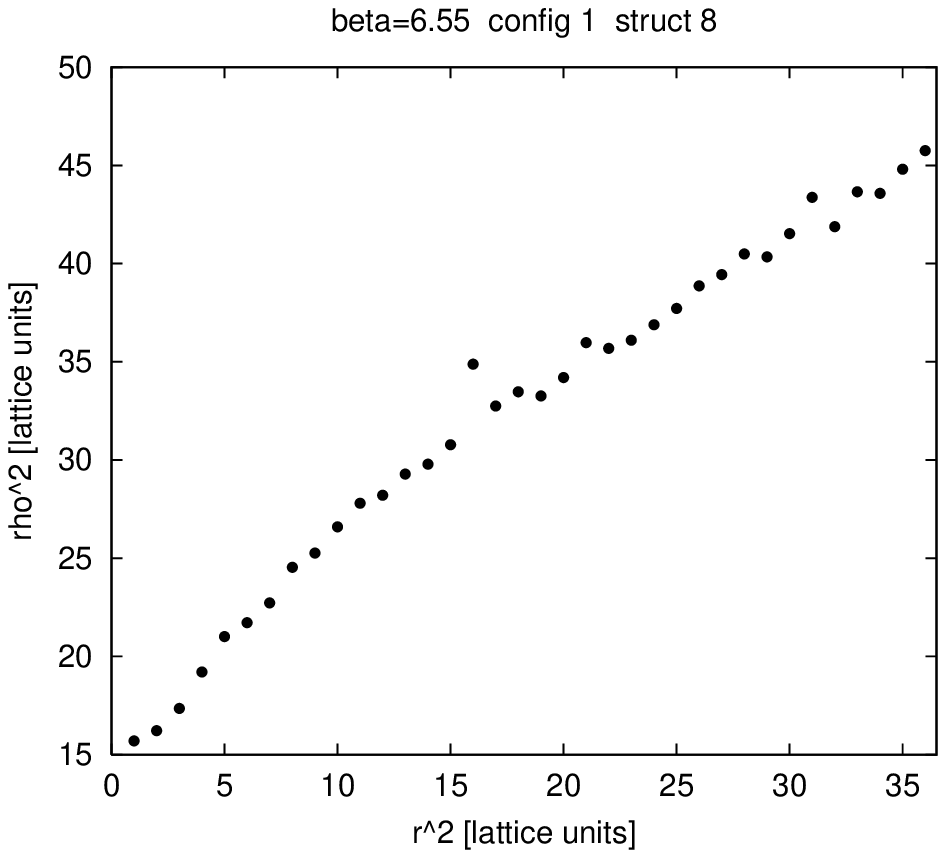}
\epsfxsize=8.0truecm\epsffile{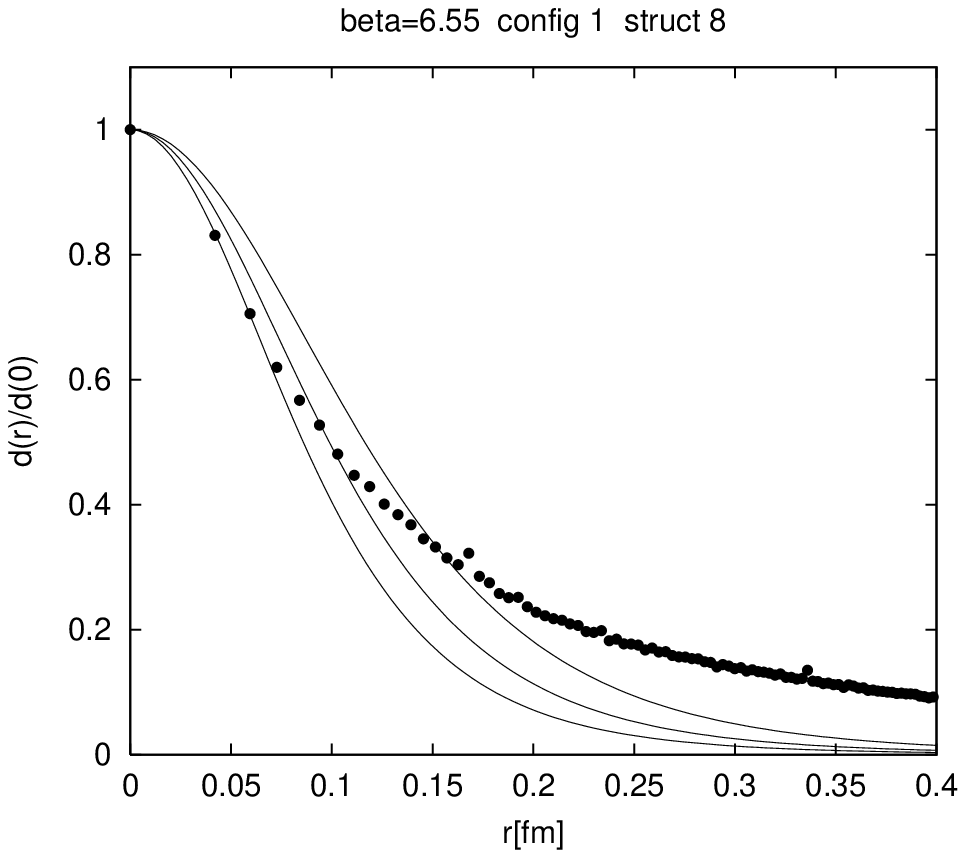}
}
\caption{Left: The function $r^2 y_n/(1-y_n)$ for a typical more intense structure.
The data should be constant for an instanton profile. Right: The profile of the 
structure and attempted fits in the region
$0.00-0.06\,\mbox{\rm fm}$ (leftmost curve), 
$0.06-0.12\,\mbox{\rm fm}$ (middle curve), and 
$0.12-0.18\,\mbox{\rm fm}$ (rightmost curve).}
\label{Instfit}
\end{center}
\end{figure}

If the near-zero mode of the Dirac operator is a mixture of ``would be'' zero modes 
associated with instantons, then peaks of the wave function should resemble the 
profile of a 't Hooft zeromode. In particular, in the ideal case we should 
have~\cite{ins_tHooft}
\begin{equation}
    {{d_n(r)} \over {d_n(0)}} \,=\, 
    \biggl( {{\rho_n^2} \over {\rho_n^2 + r^2}} \biggr)^3 \,\equiv\,
    \Bigl( y_n(r) \Bigr)^3
\end{equation} 
where $\rho_n$ is the radius of the instanton located at $n$. Thus the function,
$r^2 y_n/(1-y_n)$ is a constant ($\rho_n^2$) independent of $r$ for an instanton 
profile, and should be approximately constant for our structures if they 
represent the response to an instanton-like fluctuation. However, we find that
this is not the case and the shapes of the peaks in our low eigenmodes do not
resemble the instanton ansatz. To illustrate this, we display the situation for 
a typical structure in Fig.~\ref{Instfit}. Shown are also fits to the instanton 
profile over the the distances $0.00-0.06\,\mbox{\rm fm}$, $0.06-0.12\,\mbox{\rm fm}$ 
and $0.12-0.18\,\mbox{\rm fm}$. These fits are inconsistent with both the profile
of the peak and with each other. Note that the shapes of the peaks are already
averaged over all the directions, and should exhibit a robust behavior.

\begin{figure}
\begin{center}
\centerline{
\epsfxsize=0.7\hsize\epsffile{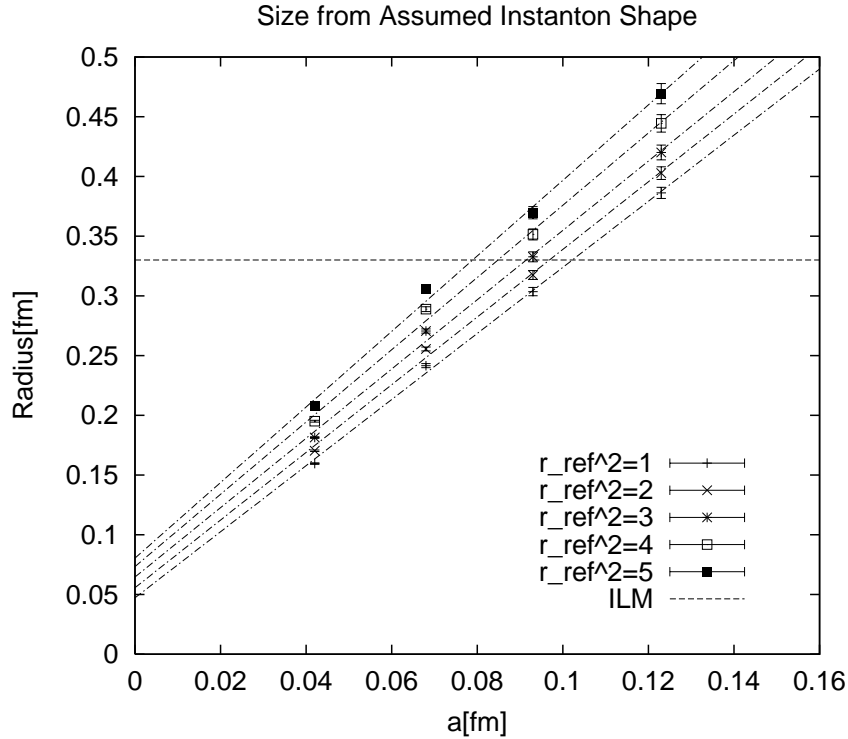}
}
\caption{Average radius, $< \rho_n >$, of structures determined from the assumed 
instanton shape of the peak at different reference distances. Lowest nonzero mode 
was used for calculation.} 
\label{Sizeinst}
\end{center}
\end{figure}

To see the inconsistency with the instanton profile on average, we choose 
a reference point $r_{ref}$ (in lattice units) and assign a radius to each 
structure through
\begin{equation}
     \rho_n(r_{ref}) \,\equiv\, 
     r_{ref}\,  \sqrt{{y_n(r_{ref})} \over {1-y_n(r_{ref})}}  
\end{equation} 
All of our structures are guaranteed to have a peaked behavior over the distance
$r_{ref}^2=3$ in lattice units, but the vast majority of them decays over much larger 
lattice distances. To treat all the structures on the same footing we compute average 
radii from the above prescription for $r_{ref}^2=1,2,\ldots,5$ and plot the resulting 
dependence on lattice spacing in Fig.~\ref{Sizeinst}. Wide vertical bands at every 
lattice spacing reflect the fact that our structures can not be fit reasonably by the 
t'Hooft profile. Nevertheless, the tendency towards sizes substantially smaller than 
the ILM value in the continuum limit is still obvious, thus showing that the overall 
conclusion from Fig.~\ref{Sizechir} remains valid even for this definition of the 
radius.

We emphasize that contrary to results in Fig.~\ref{Sizechir}, we do not assign 
much physical significance to the precise behavior of data in Fig.~\ref{Sizeinst}. 
The main purpose of showing these results is to reveal the marked inconsistencies 
with the instanton picture of the vacuum.

\subsection{Correlation Functions of Local Chirality}

We have also computed the chirality-chirality correlation function in the lowest 
Dirac nonzero mode $\psi$, namely
\begin{equation}
       C_{cc}(r) \;\equiv \; \langle\, c(n) c(m)\, \rangle_{|n-m|=r}\, V^2
\end{equation}   
This correlator has been studied in Ref.~\cite{deGrand00A}. While the average
information encoded in the correlator may be useful, it will not provide us with
a detailed view of the space-time distribution of $c(n)$. The point is that 
if this $C_{cc}(r)$ has characteristic size $\rho$, one still cannot infer whether 
the typical space-time distribution of $c(n)$ in a configuration predominantly 
appears localized on spherical structures of radius $\rho$, or if it comes in one 
of the infinitely many other forms leading to the same average correlation. 

\begin{table}[t]
  \centering
  \begin{tabular}{ccccc}
  \hline\hline\\[-0.4cm]
  \multicolumn{1}{c}{$a$ [fm]}  &
  \multicolumn{1}{c}{0.123}  &
  \multicolumn{1}{c}{0.093}  &
  \multicolumn{1}{c}{0.068}  &
  \multicolumn{1}{c}{0.042}  \\[2pt]
  \hline\\[-0.42cm]
   $<r^2>^{1/2}$ [fm] & 0.33 & 0.24 & 0.22 & 0.24 \\
   $<r>\quad\;\;$ [fm]   & 0.26 & 0.19 & 0.17 & 0.19 \\
  \hline \hline
  \end{tabular}
\caption{Estimates of the size for the correlator $C_{cc}(r)$.}
\label{Sicor_est} 
\end{table}

\begin{figure}
\begin{center}
\centerline{
\epsfxsize=0.7\hsize\epsffile{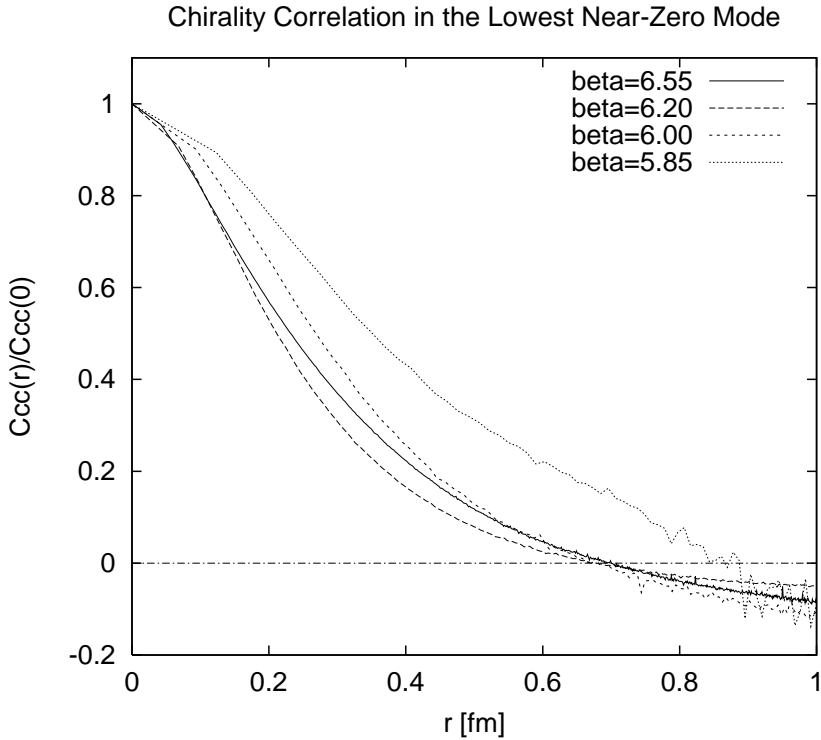}
}
\caption{Average $C_{cc}(r)/C_{cc}(0)$ from lowest near-zero mode.} 
\label{Corfs}
\end{center}
\end{figure}

The average $C_{cc}(r)/C_{cc}(0)$ from the lowest nonzero modes for our four ensembles 
are shown in Fig.~\ref{Corfs}. While larger statistics would be desirable, we have 
estimated the ``size'' of the correlators by evaluating $<r>$ and $<r^2>^{1/2}$ with 
$C_{cc}(r)$ used as a probability distribution. In every case we have cut off 
the integration at the distance where the correlator first turns negative. 
The resulting values are collected in Table~\ref{Sicor_est}. Note that values
for $<r^2>^{1/2}$ are systematically higher than sizes obtained from coherent local 
chirality (c.f. Fig.~\ref{Sizechir}) which is to be expected for an inhomogeneous 
space-time distribution of fluctuations with various shapes. Nevertheless, the sizes 
of correlators we have obtained are still systematically lower than the ILM value 
for instanton radius.

\subsection{The Interpretation of the Results}
\label{subsec:interpret}

The data in Figures~\ref{density} and~\ref{Sizechir} represent rather interesting
new results and we would now like to elaborate on their interpretation. We have 
started from the assumption that the vacuum has local properties identifiable with
ILM scenario, 
i.e. that it typically fluctuates in such a way as to form (anti)self-dual 
lumps of approximately quantized topological charge. However, the identification 
of such presumed lumps resulted in quantitative characteristics that are in 
marked disagreement with the ILM on fine lattices. We observe many more 
structures of much smaller physical size (although this size remains 
{\it finite} even in the continuum limit).
Given that the parameters of the ILM are rather tightly fixed~\cite{ILM},
the true microscopic picture as seen by the lattice fermion appears to be 
very different from that envisioned by the ILM.

Particularly puzzling from the instanton perspective is a large number of peaks 
that we observe in the lowest near-zero modes, implying that a propagating 
light quark feels many more ``kicks'' from the regions of strong gauge field 
than one would expect based on the ILM. Indeed, assuming that the underlying gauge 
structures represent elementary instanton tunneling events between classical 
vacua leads to clear contradictions. For example from our results at $\beta=6.55$ 
one would estimate that the gluon condensate should be of the order 
$<0|G^2|0>\approx 32\pi^2\,n \approx 32\pi^2\, 50\, \mbox{\rm fm}^{-4} \approx 
25$ GeV$^4$, where $32\pi^2$ is a contribution of a single 
instanton \footnote{The estimates of this type are usually used to justify
the ILM parameters~\cite{ILM}.}. This is to be compared with the accepted value 
$<0|G^2|0>\approx 0.5-1.0$ GeV$^4$. Similar inconsistency arises from considering 
topological susceptibility under even weaker assumptions: Assuming that gauge 
structures are general lumps of quantized and approximately uncorrelated 
topological charge leads to the estimate of topological susceptibility
$<Q^2>/V \approx n \approx 50\, \mbox{\rm fm}^{-4} \approx 
50 (200 \mbox{\rm MeV})^4$, i.e. about fifty times the accepted value.
At the same time, the topological charges of our lattices (determined as
a byproduct of the overlap calculation) are in very good agreement with 
the expected value of topological susceptibility in the pure gauge
vacuum~\footnote{For example, from our small ensemble at $\beta=6.55$ we  
estimate the topological susceptibility $\chi=0.92 \pm 0.50$ fm$^{-4}$.}.

\begin{figure}
\begin{center}
\centerline{
\epsfxsize=0.7\hsize\epsffile{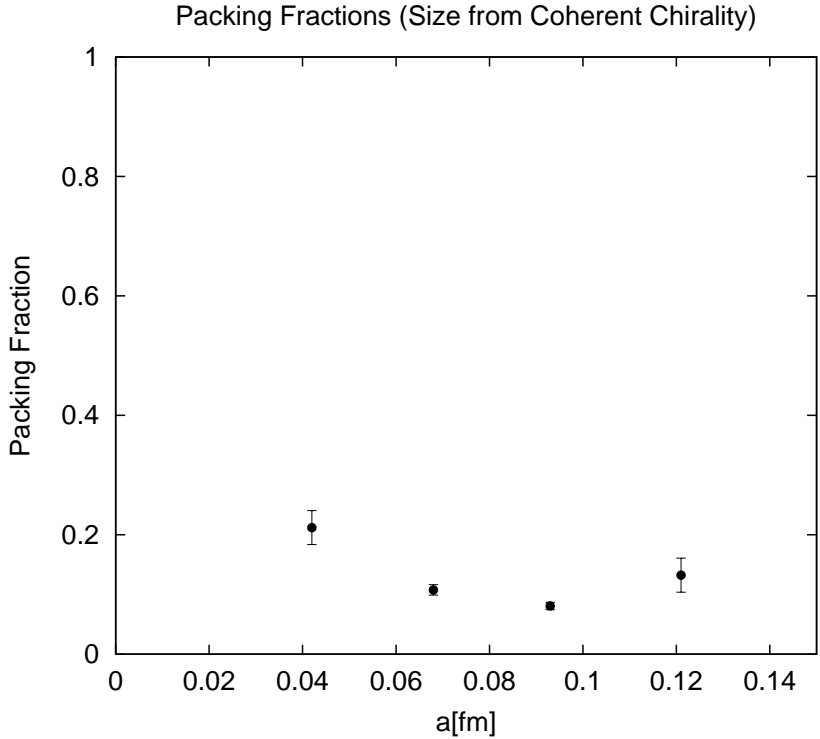}
}
\caption{Packing fractions. The sites contained in a given structure are bound 
by the largest hypersphere with the same sign of local chirality as in the 
center of the structure.}
\label{Packings}
\end{center}
\end{figure}

The above considerations strongly suggest that the topological charge contained 
within the structure is probably not quantized. One is led to the same 
conclusion from consideration of radii of regions of coherent local chirality. 
Our data indicate the value $< R_n > \approx 0.15\,\mbox{\rm fm}$ in the continuum 
limit. However, the 't Hooft instanton formula for density of quantized charges 
valid at short distances, i.e. $n(\rho)\propto \rho^6$, is frequently invoked as 
a basis for concluding that the occurrence of instantons of this size should be 
essentially negligible (certainly relative to $\rho\approx 0.3\, \mbox{\rm fm}$ 
instantons for which the formula is still expected to be valid)~\cite{Teprev,ILM}. 
We interpret this marked mismatch as a manifestation of the fact that the 
underlying assumption for the 't Hooft formula, namely the local quantization 
of topological charge, is probably not valid. Indeed, the suppression of small 
instantons is a consequence of their quantized topological charge: it is difficult 
to squeeze an entire unit of topological charge inside a small radius. However, 
without quantization, there is no reason to expect a suppression of smaller-scale 
fluctuations, which would simply contain proportionally less topological charge.

Finally, we would like to return to Fig.~\ref{density} and discuss the rapidly 
increasing density of the structures with decreasing lattice spacing. While the
density at $\beta=6.55$ is very large, the fraction of sites contained in the 
structures is still relatively small as shown by the packing fractions plotted
in Fig.~\ref{Packings}. The peaks thus retain their identity as can be seen from 
Fig.~\ref{Structures}. However, let us elaborate for the moment on the hypothetical
possibility that the density increases indefinitely, because this might suggest 
something unphysical even if topological charge is not locally quantized as we
argued above. Indeed, how should one interpret a diverging density of structures
in the continuum limit, especially if their average size (c.f. Fig.~\ref{Sizechir}) 
remains finite? The point is that contrary to quantized topological charges, where 
the finite topological susceptibility forces the density to assume a finite physical 
value in the continuum limit, the peaks of non-quantized topological charge might 
well lose their identity closer to the continuum limit and their density be indeed 
unphysical. (However, the regions of coherent local chirality identified with the help 
of these peaks, and their sizes are still physical.) The picture that we have in mind 
is that of relatively isolated mountain peaks belonging to a larger mountain range
and coalescing as the continuum limit is approached. Our algorithm to measure the size 
of regions of coherent local chirality would then sample the size of the mountain 
range rather than individual peaks. A more dynamical analogy is to imagine disturbing 
a calm surface of water. The regions where the water is above the original level 
(positive topological charge density) have various shapes and volumes, but on average 
have some typical size. Yet, within such regions there could be many local maxima at 
all length scales. While the sizes of individual maxima do not define the scale and 
may not be physically meaningful, the average radius measured relative to these maxima 
probes the typical finite size of these coherent regions. It is easy to imagine that 
in the continuum, the fluctuations of topological charge resemble violently disturbed 
surface of water. In such a hypothetical scenario, the diverging density of structures 
would actually be very natural.

We emphasize that in this paper we do not attempt to put forward any specific 
low-energy scenario for the behavior of topological charge fluctuations in the continuum 
limit. The purpose of the discussion in the previous paragraph is to argue that 
as soon as one abandons the local quantization of topological charge, then the data 
in Fig.~\ref{density} do not represent anything unexpected.

\subsection{The Question of Dislocations}

In the lattice discussions of gauge field topology one has to face the possible 
problem of ``dislocations''. While this notion is frequently used as an unspecified 
synonym for ``lattice artifact'', in the original discussion of 
Ref.~\cite{sdisloc,gdisloc} it has a rather well defined meaning which we will 
adopt. In particular, for a given lattice gauge action and given topological charge 
operator, it is the local structure in the gauge field with action smaller 
than the continuum action of an instanton, and with unit topological 
charge~\footnote{We thank Tam\'as Kov\'acs for clarifying this.}. In the 
loose sense, this is frequently pictured as a small instanton that is almost 
falling through the lattice and lives on 1-2 lattice spacings. Invoking 
entropy arguments, it was suggested in Ref.~\cite{gdisloc} that for the 
action--topological charge combination where this happens, it is possible that 
dislocations dominate the topological charge fluctuations, and can lead 
to unphysically large (possibly diverging) susceptibility. This raises 
the question of whether the large density of structures that we see is caused
by unphysical dislocations~\footnote{We thank Thomas Sch\"afer for
pointing this out.}.

\begin{figure}
\begin{center}
\centerline{
\epsfxsize=8.0truecm\epsffile{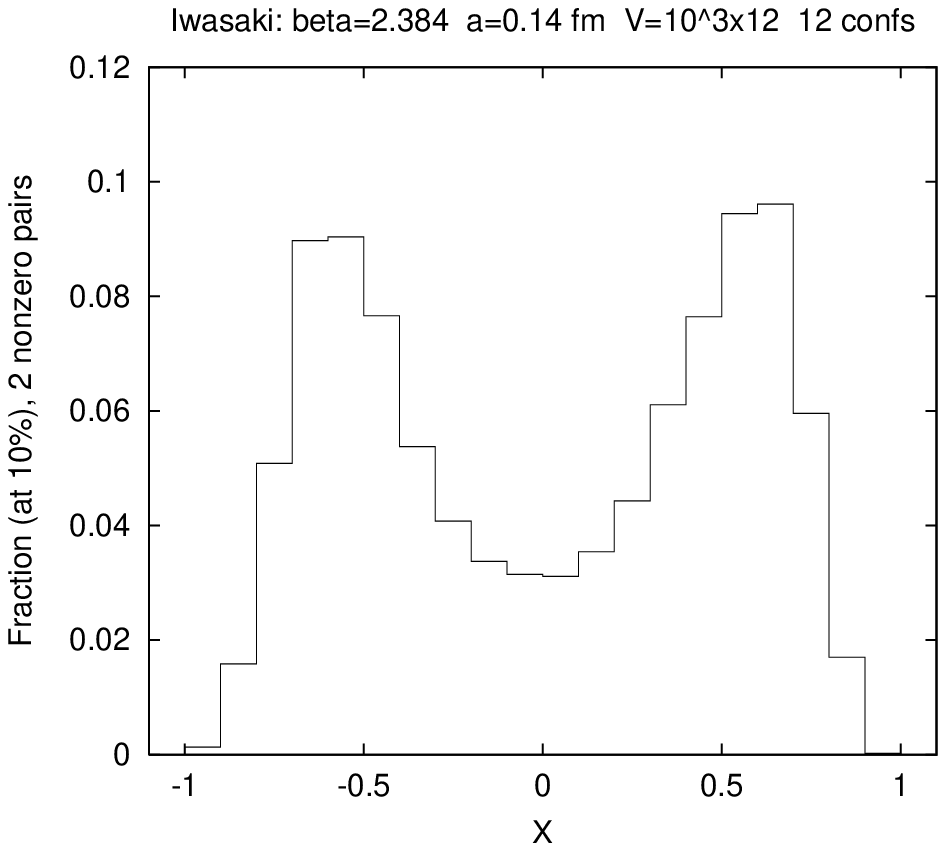}
\epsfxsize=8.0truecm\epsffile{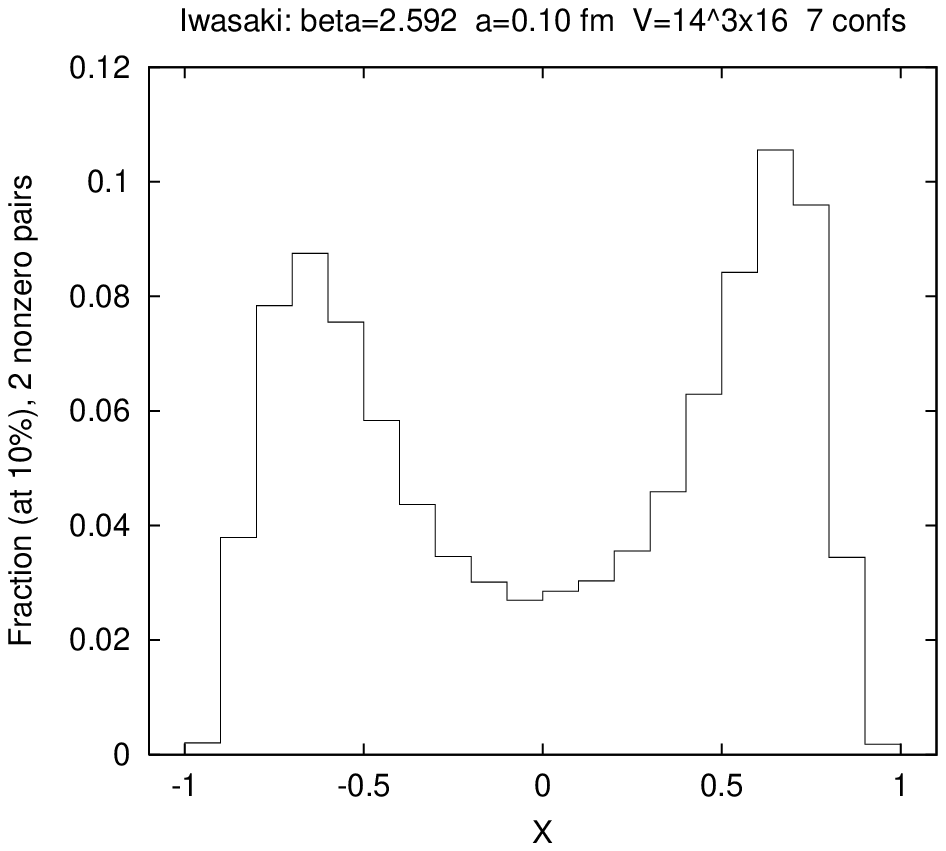}
}
\vskip 0.15in
\centerline{
\epsfxsize=8.0truecm\epsffile{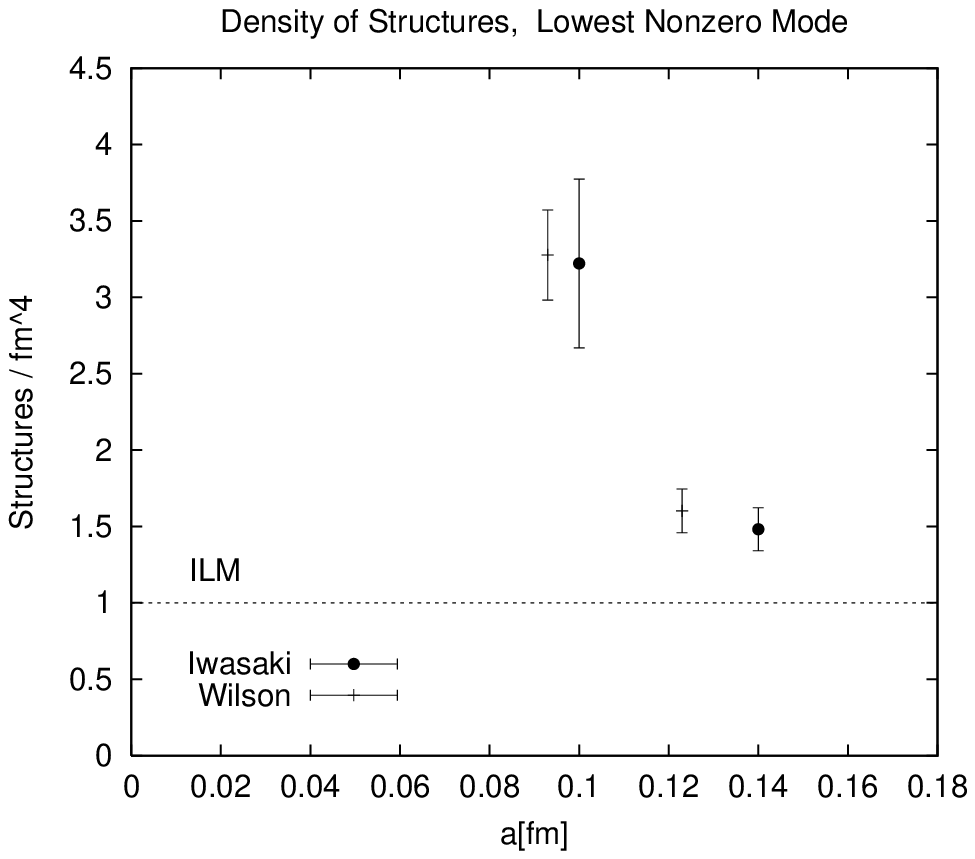}
\epsfxsize=8.0truecm\epsffile{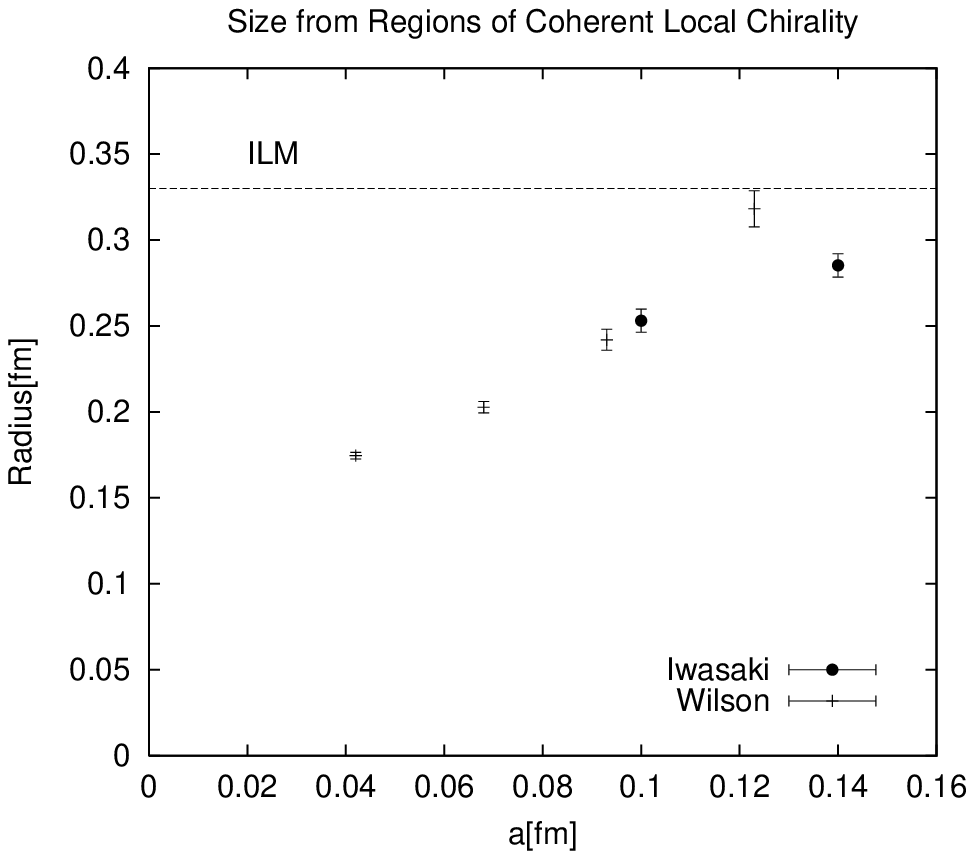}
}
\caption{Upper: $X$-distributions for two Iwasaki gauge ensembles considered.
         Lower left: Density of structures for Iwasaki and Wilson actions. 
         Lower right: Average size from coherent regions of local chirality
         for Iwasaki and Wilson actions.}
\label{Iwasaki}
\end{center}
\end{figure}

This question can be addressed at several levels. First of all, the 
arguments of Ref.~\cite{gdisloc} assume that the topological charge
of the dislocation is concentrated on its core. Indeed, the idea that
dislocations are important relies rather heavily on the picture of locally
quantized topological charge. If the topological charge does not come in unit 
lumps, as we suggest here, then the issue simply does not arise. 

Secondly, it is important to realize that dislocation is a concept 
assigned to the pair gauge action--topological charge operator. While
we work with Wilson gauge action which is supposedly susceptible to the 
possibility of dislocations, we use chiral fermions to measure topological 
charge. It has frequently been noted (see e.g. Ref.~\cite{ILM}), that fermions 
should not be sensitive to dislocations. This is indeed very plausible. As we 
have argued in Sec.~\ref{subsec:fermions}, one naturally expects that infrared 
fermionic modes will be quite smooth and will not inherit the singular behavior 
of the underlying gauge fields.

Finally, there is no hint of behavior symptomatic of dislocations in our data. 
For example, even the smallest of our structures are extended objects 
in physical units and span at least 4 lattice spacings. Moreover,
the estimates of topological susceptibility from topological charge measured
by the overlap Dirac operator are in very good agreement with the accepted value 
of about $1\,\mbox{\rm fm}^{-4}$ (see footnote 8). If the underlying structures 
were dislocations (and hence carried topological charge close to unity), then 
this value should be substantially larger. This is one of the arguments that 
actually suggests that topological charge is not locally quantized, as we 
stressed in Sec.~\ref{subsec:interpret}.

Nevertheless, to eliminate any concern that dislocations may be a problem, we 
have performed a comparative study with a renormalization group improved gauge 
action. In this case, the possibility of dislocation dominance should be 
substantially reduced. In particular, we have employed the Iwasaki 
action~\cite{Iwasaki}, and studied overlap eigenmodes at two different lattice 
spacings. In the standard normalization for Iwasaki action, 
(see Refs.~\cite{Iwasaki,appliwasaki}), we considered a 10$^3$x12 lattice
at $\beta=2.384$ and a 14$^3$x16 lattice at $\beta=2.592$. This corresponds
to lattice spacings $a=0.14\,\mbox{\rm fm}$ and $a=0.10\,\mbox{\rm fm}$
respectively, as determined from string tension by using the data of
Ref.~\cite{appliwasaki} and the two-loop beta function fit.

The summary of our results is shown in Fig.~\ref{Iwasaki}. The structure of 
the resulting $X$-distributions is very similar to the structure for Wilson 
gauge ensembles as can be seen by comparing to Fig.~\ref{Xdists}. We thus applied
identical procedures (with identical cuts) for finding the structures and 
determining their size as we did for Wilson gauge ensembles. Although the behavior
of density for Wilson gauge action (see Fig.~\ref{density}) might have suggested 
the possibility of dislocations, this is not the case because a similar pattern
appears to be followed in the case of Iwasaki action as well. Similarly,
the sizes of structures are systematically lower than the ILM values. 
It would obviously be interesting to investigate a larger window of lattice 
spacings with Iwasaki action, to be able to study the issues of continuum 
extrapolation. However, the purpose of this comparative study was to demonstrate 
that our conclusions are not sensitive to the choice of gauge action and that
there is no problem of dislocations. Our results clearly indicate that this is
indeed the case.

\section{Conclusions}

Uncovering the local structure of topological charge fluctuations promises
to have profound implications for our understanding of low energy QCD. This
expectation is born out of the fact that possible microscopic explanations for 
important phenomena such as spontaneous chiral symmetry breaking, the resolution 
of the U$_A$(1) problem, and the $\Theta$-dependence of QCD, are based on the 
picture of the vacuum wherein self-dual lumps of locally quantized topological
charge (instantons) play a major role. It is consequently quite essential
to determine whether such picture (e.g. ILM) is indeed fundamental, or if there is in 
fact a different microscopic mechanism driving these important effects.

Following upon the ideas of Ref.~\cite{Hor01A}, we have explained here in detail 
why studying low-lying Dirac eigenmodes provides a natural and reliable approach 
for exploring the local nature of topological charge fluctuations (at least as
implied by the ILM). The study of the $X$-distribution in low-lying modes has 
been designed as a tool for probing the local vacuum structure indirectly. 
While the available results~\cite{Followup,Gattr01A} on qualitative behavior 
of $X$-distribution could in principle be viewed as confirming the consistency 
of the instanton picture, we have pointed out here that the association with 
instantons is not unique. In other words, the double-peaked qualitative 
structure of $X$-distribution is a necessary condition for instanton dominance, 
but not sufficient.

\begin{figure}
\begin{center}
\centerline{
\epsfxsize=5.4truecm\epsffile{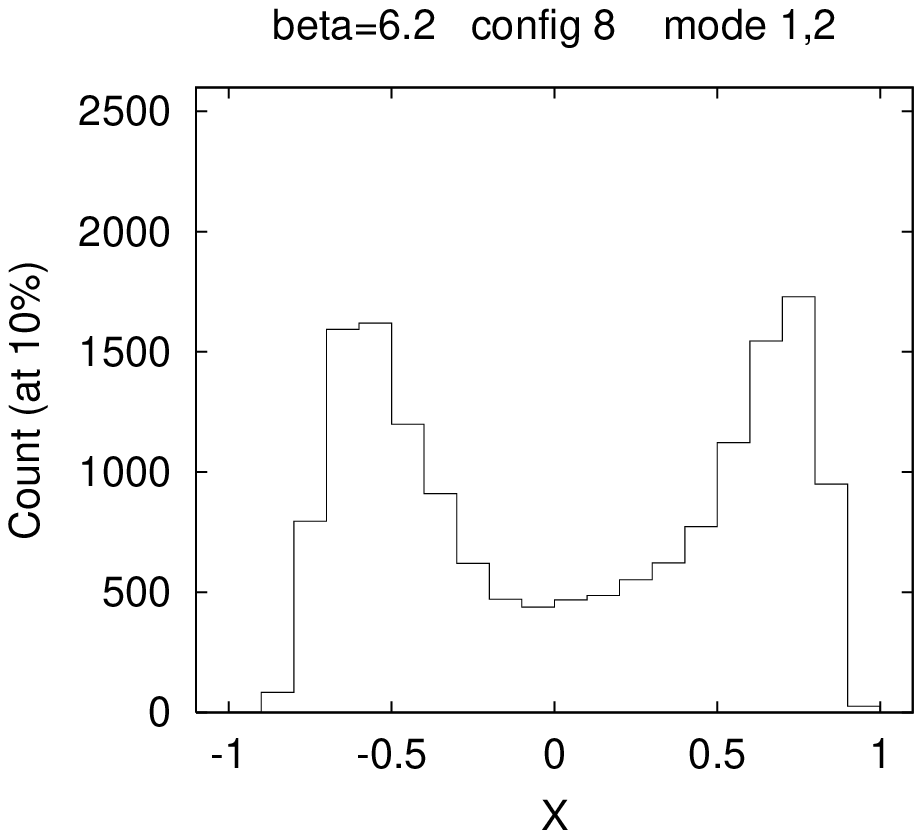}
\epsfxsize=5.4truecm\epsffile{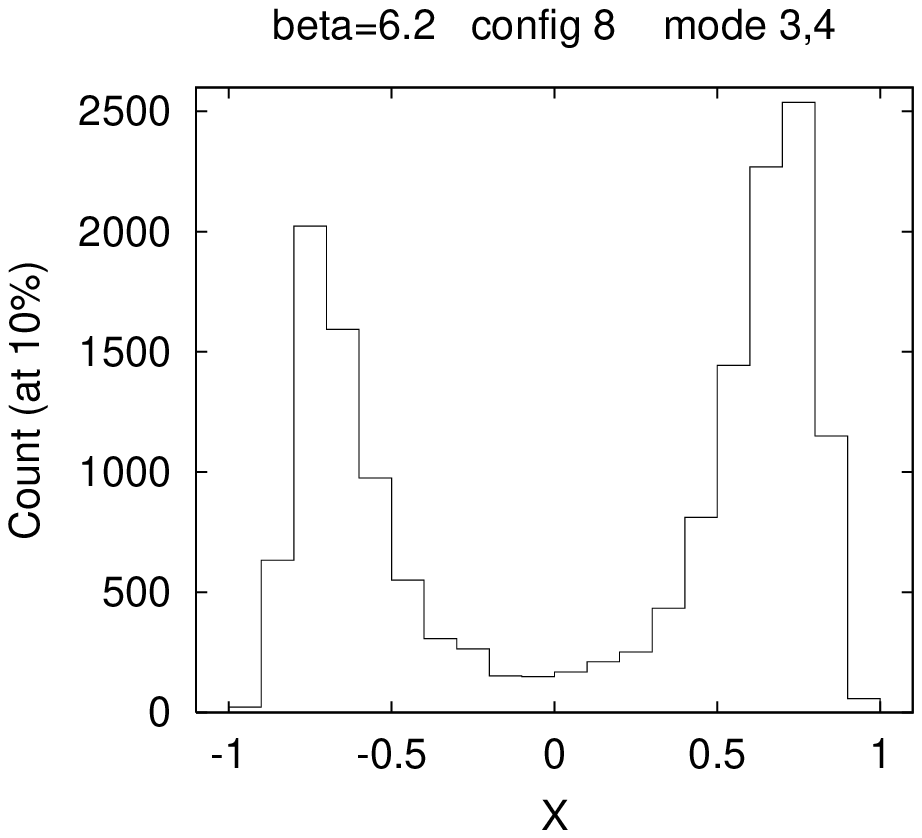}
\epsfxsize=5.4truecm\epsffile{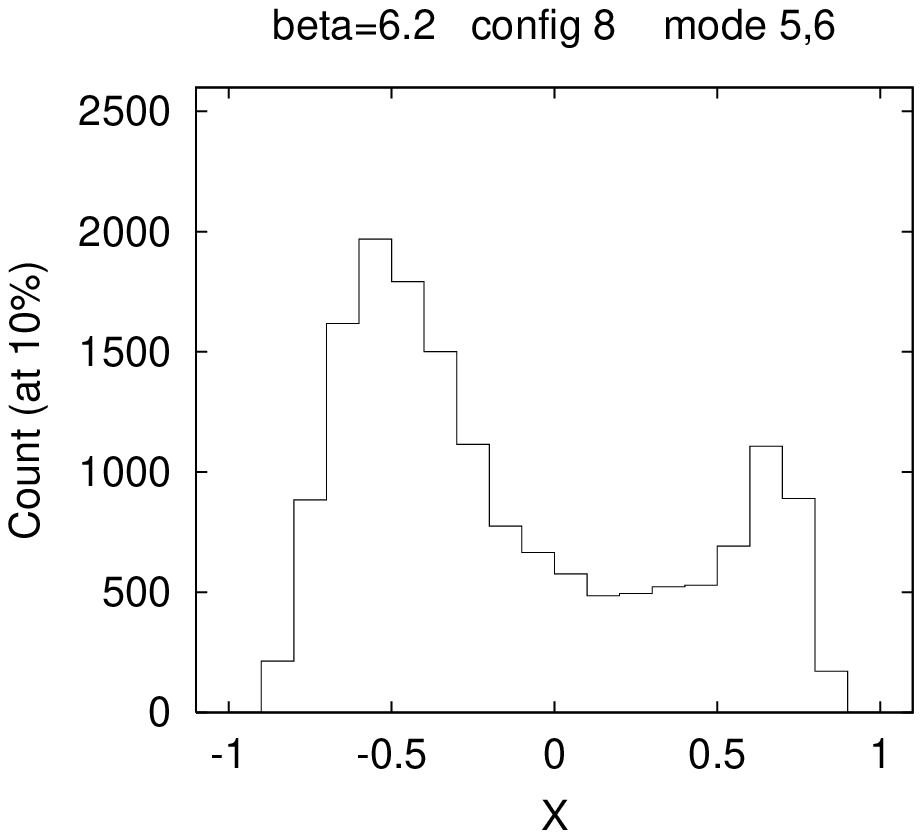}
}
\vskip 0.15in
\centerline{
\epsfxsize=5.4truecm\epsffile{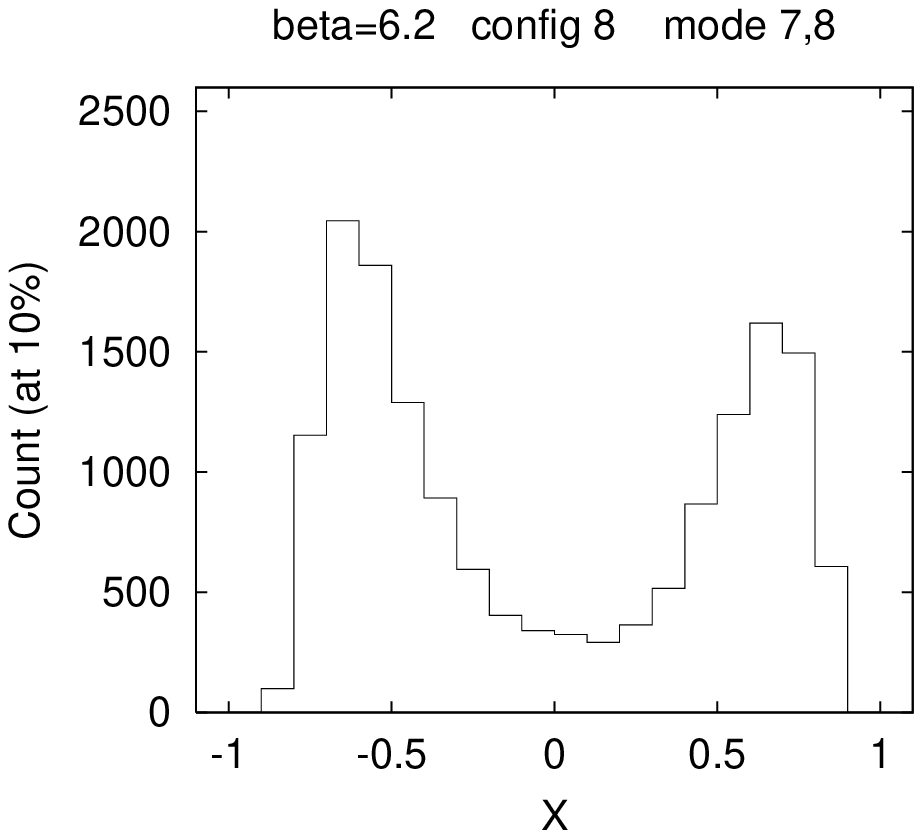}
\epsfxsize=5.4truecm\epsffile{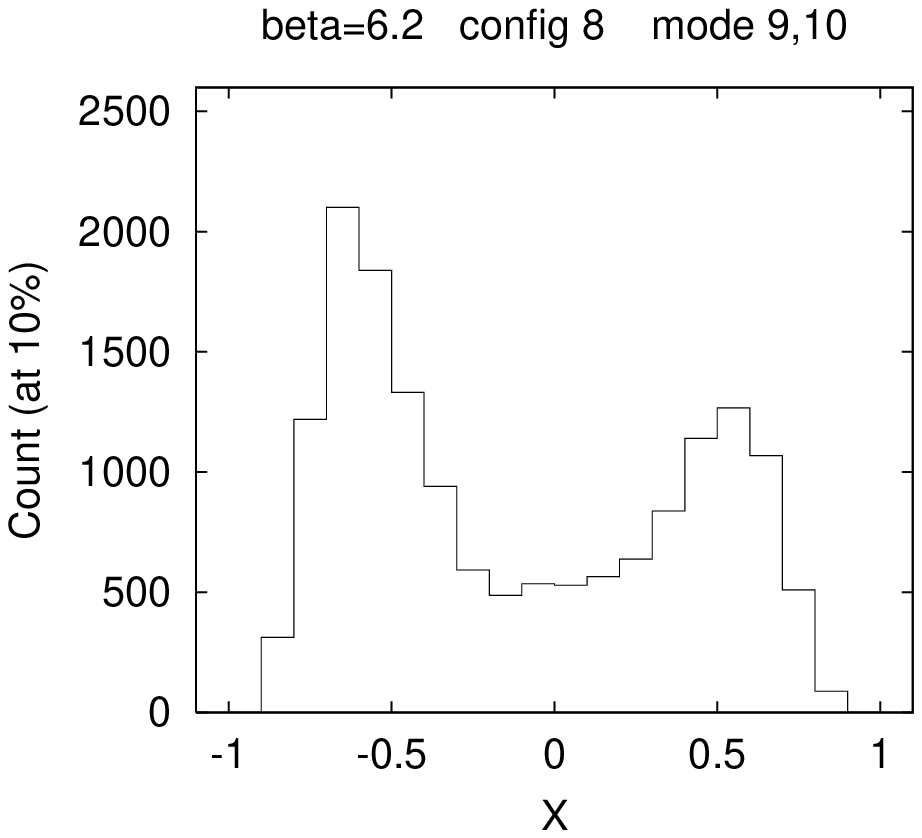}
\epsfxsize=5.4truecm\epsffile{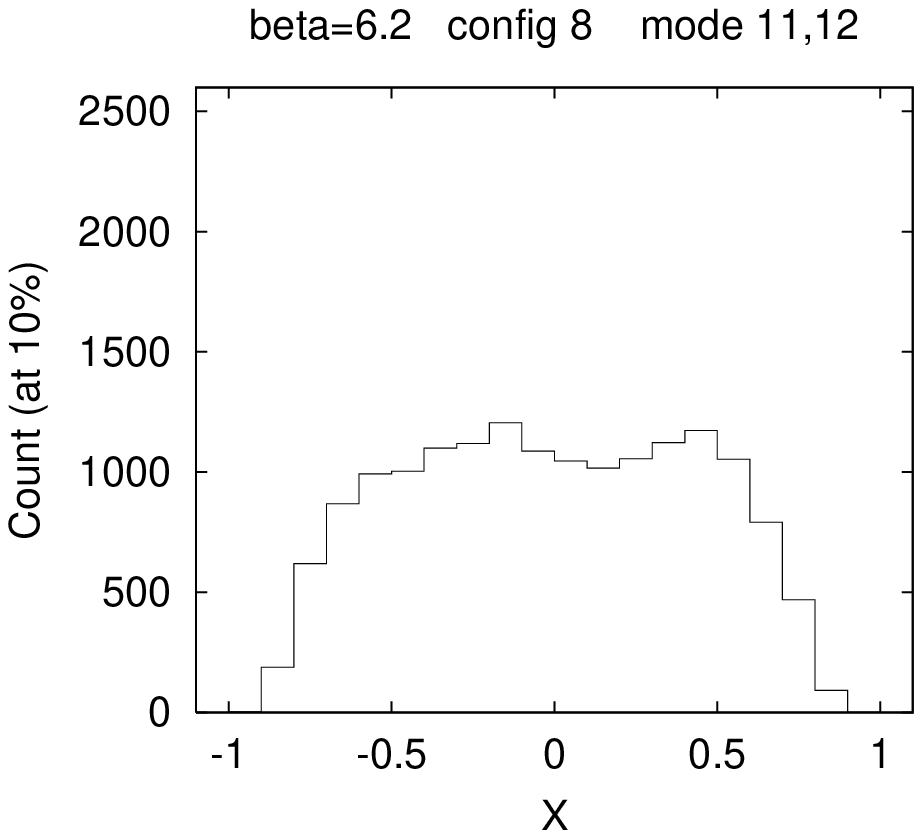}
}
\vskip 0.15in
\centerline{
\epsfxsize=5.4truecm\epsffile{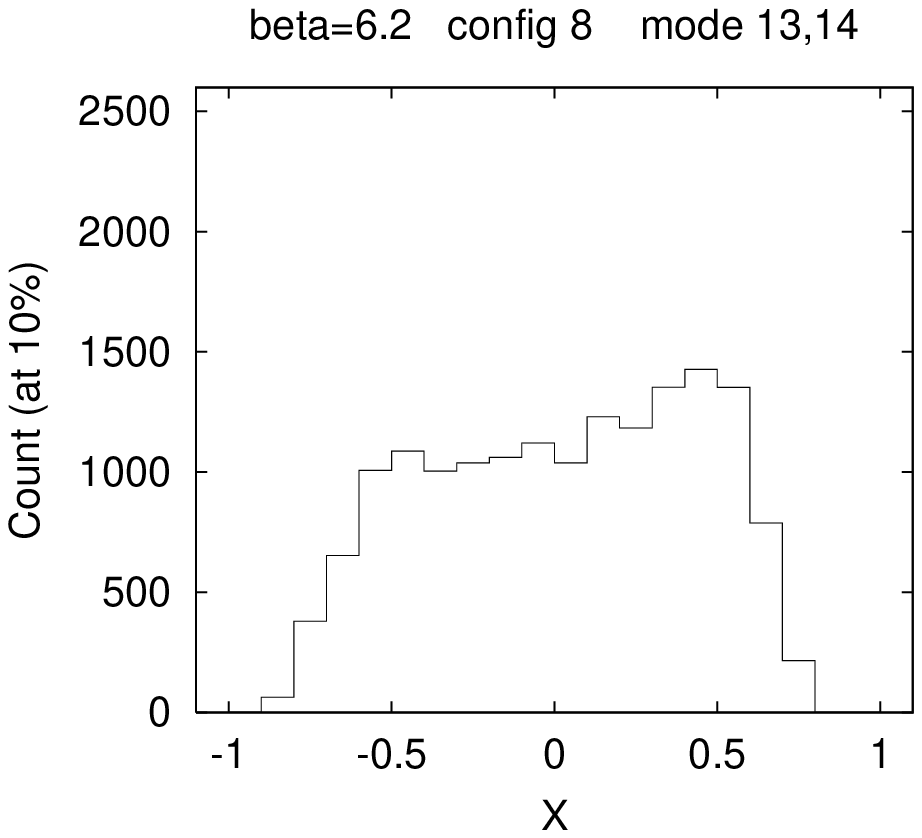}
\epsfxsize=5.4truecm\epsffile{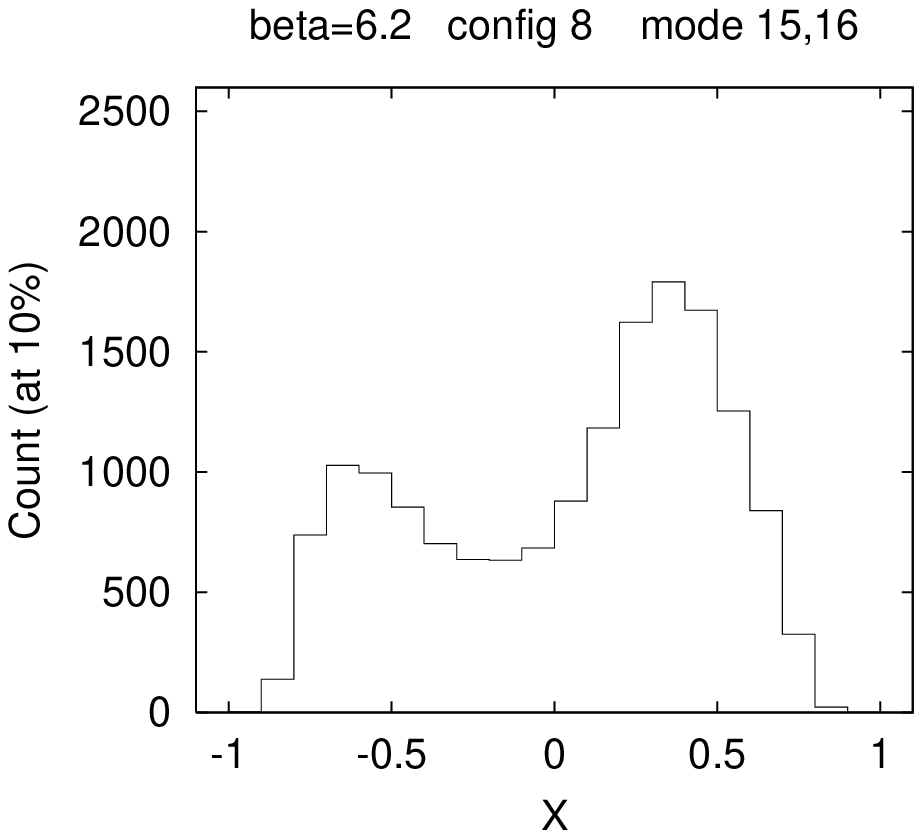}
\epsfxsize=5.4truecm\epsffile{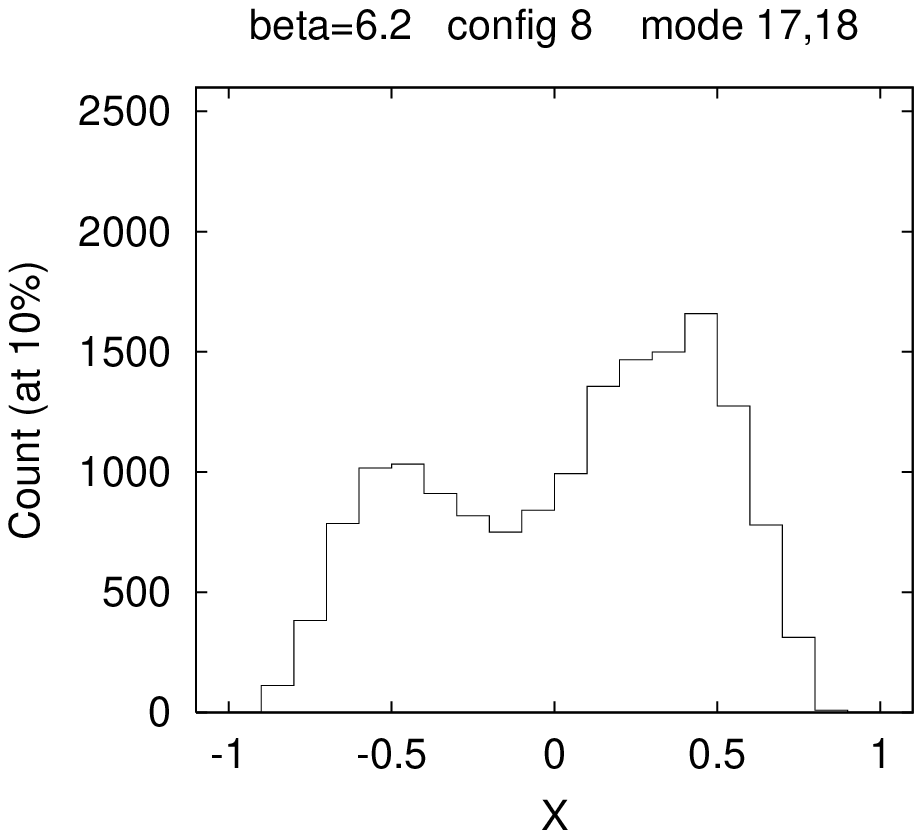}
}
\vskip 0.15in
\centerline{
\epsfxsize=5.4truecm\epsffile{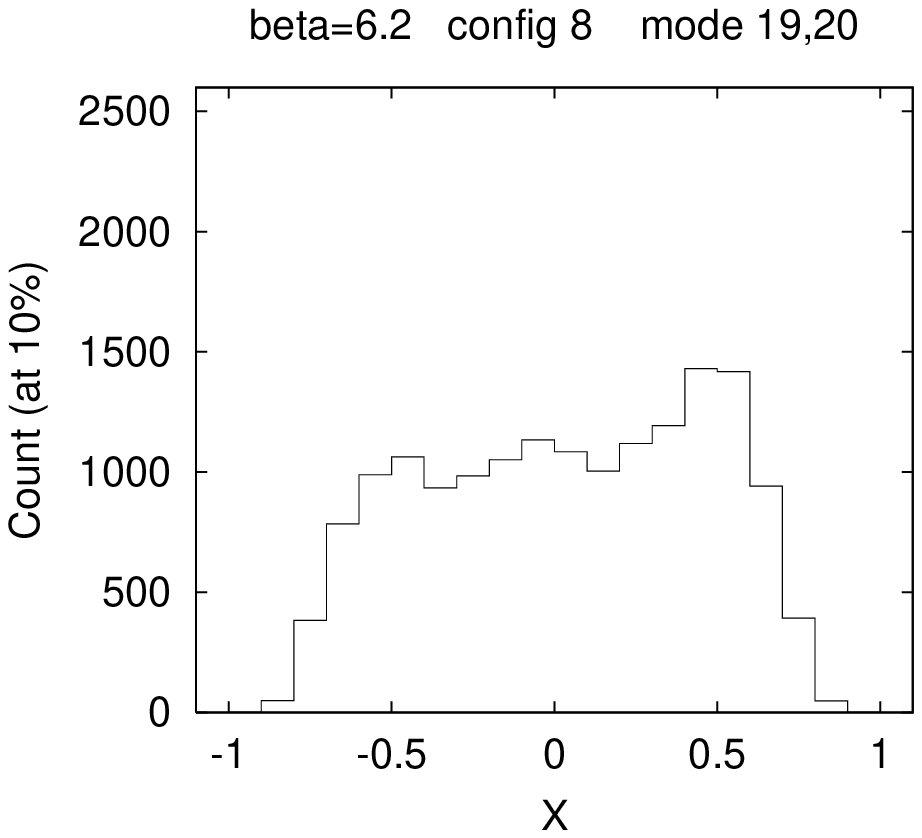}
}

\caption{The $X$-distributions for first 20 near-zero modes of configuration 8
at $\beta=6.2$.}
\label{Killer}
\end{center}
\end{figure}

As a step towards resolving these issues, we have computed and analyzed 
low-lying modes of the overlap Dirac operator in Wilson gauge backgrounds 
over a wide range of lattice spacings ($a\approx 0.04-0.12\,\mbox{\rm fm}$).
The double-peaked structure of the $X$-distribution has been observed 
with maxima at $X\approx \pm 0.65$ (using a fraction $f=0.1$ of lattice points).
The position of maxima shows very little change with lattice spacing without 
visible tendency to move toward the extremal values of $X=\pm 1$, even though 
an exactly chiral lattice fermion is used. Our main initial aim in this study was 
to verify whether the characteristics of local structures contributing to 
the peaks of the $X$-distribution are consistent with quantitative characteristics 
of the gauge features in the backgrounds proposed in the ILM. In other words,
we attempted to verify whether low-lying modes can (at least approximately)
be viewed as mixtures of t'Hooft ``would be'' zeromodes generated by
gauge structures resembling approximately $\rho=\third$ fm instantons of 
density $n=1$ fm$^{-4}$. To the contrary, we have found local characteristics 
indicating that the true nature of topological charge fluctuations is very 
different from that envisioned in the ILM. 

We emphasize in this context that the underlying issue here is to determine 
whether one should view the ILM picture as a fundamental one, or as 
an effective description with phenomenological meaning only. For example, 
claims that the bulk of spontaneous chiral symmetry breaking is due 
to mixing of ``would-be'' zeromodes associated with instantons have only well 
defined microscopic meaning if the corresponding 
low-lying modes have the local structure predicted by this microscopic mechanism. 
To verify this proposition, we have no other choice but to examine these modes. 
This appears to be a well-defined problem with a well-defined answer. Our results 
indicate that the ILM scenario does not provide for accurate microscopic 
description of these modes and thus remains at the phenomenological level. It is 
this important distinction that we wish to stress in this paper.

While the detailed arguments are summarized in Sec.~\ref{subsec:interpret}, 
our main conclusion can be verified in an independent manner. To make this point 
qualitatively, we have calculated twenty near-zeromodes for configuration 8 from 
our $\beta=6.2$ ensemble~\footnote{Needless to say, this was rather demanding on 
computer resources.}. This configuration has $Q=0$, and the physical volume is such 
that according to the ILM it should contain about 3 or 4 instantons and antiinstantons. 
Hence, the subspace spanned by t'Hooft would-be zeromodes should have a dimension of 
that order. Consequently, it is the ILM prediction that chirally peaked $X$-distributions 
should be observed for 3 or 4 near-zeromodes (and possibly a few more) but the rest of 
the modes should resemble approximate free-field behavior with local chirality 
peaked around the origin. In Fig.~\ref{Killer} we plot the $X$-distribution 
at $f=0.1$ for all 10 pairs of near-zeromodes computed (the histogram is the same 
for both modes in a pair). Inspecting these results reveals that there are at least 
14 modes with significant double-peaked structure and none of them is peaked at 
the origin. As a matter of fact, at $f=0.02$ {\it all} 20 calculated modes exhibit 
the double-peaked behavior and this most likely persists even for higher modes. 
While this fact can hardly be explained by the ILM, it is not very surprising in 
view of the results presented here. On this particular configuration, we have identified 
32 local structures.~\footnote{Note that this observation makes also very unlikely 
the possibility that one could interpret several structures as forming a severely 
deformed quantized lump of topological charge which one could somehow still 
associate with the ILM instanton. In this case the subspace of would-be zeromodes 
should still be of size 3 or 4.}

Even though we have concentrated on the ILM in this paper, we believe that our results
suggest a more general conclusion. In particular, as we have argued in
Sec.\ref{subsec:interpret}, it is very difficult (if not impossible) to reconcile 
the topological susceptibility of the pure gauge vacuum and the assumption that 
local structures in fermionic near-zeromodes are caused by underlying gauge excitations 
with locally quantized topological charge. This leads us to believe that the bulk
of topological charge in the QCD vacuum is not locally quantized in integer units 
as suggested some time ago by Witten~\cite{WittenUA(1)}. 
We will address this issue in detail in a forthcoming publication~\cite{quantiz_paper}. 

\bigskip\medskip
\noindent
{\bf Acknowledgments:} 
This work has been supported in part by U.S. Department of Energy under grants
DE-FG05-84ER40154, DE-FG02-95ER40907, DE-FG02-97ER41027 and DE-AC05-84ER40150.
We have benefited from discussions with Philippe deForcrand and Tam\'as Kov\'acs.

\end{document}
\bye